# Plasma simulations with adaptive mesh and algorithm refinement

Vladimir I. Kolobov[a*] and Robert R. Arslanbekov[b]

[a] The University of Alabama in Huntsville, Huntsville, AL 35899, USA

[b] CFD Research Corporation, Huntsville, AL 35806, USA

[*] Corresponding author: vik0001@uah.edu

## Abstract

Adaptive Mesh and Algorithm Refinement (AMAR) combines adaptive Cartesian meshes with automatic selection of kinetic or fluid solvers across a computational domain. It was introduced for rarefied gas dynamics in the Unified Flow Solver and extended to gas mixtures, particle and lattice-Boltzmann solvers, computational electromagnetics, multi-phase flows, and heterogeneous CPU–GPU architectures. In weakly ionized plasmas, the species differ by orders of magnitude in mass, velocity, and time scale, which was not the case in gas-dynamic applications. This paper describes PAMAR, a plasma extension of AMAR in which electrons, ions, neutral species, and electromagnetics can have independent adaptive-mesh solvers with their own phase-space dimensionality, refinement criteria, and time step, coupled through a field solver in one executable. We present two demonstrations: a transient simulation of a planar capacitively coupled plasma (CCP) in argon using 1d1v two-stream kinetic solvers for electrons and ions on independent adaptive quadtrees, and a steady spherical plasma problem using 1d2v kinetic solvers. The CCP simulations resolve three electron groups and the different time scales of electron and ion kinetics without statistical noise, balancing cost and detail between particle-in-cell (PIC) simulations and fast CCP models. A sphere at floating potential in a collisionless or weakly collisional plasma is computed with 1d2v $(r, v, \mu)$ octree solvers for both species and compared with an exact method-of-characteristics (MoC) solution. Per-species meshes resolve the cold ions at a tenth of the uniform-mesh cost, and graded meshes improve flux conservation, collected currents, and electron density profiles. Both demonstrations also select the model cell by cell. In the CCP, a hybrid ion solver is kinetic in the sheaths and fluid in the bulk. Around the sphere, a hybrid electron solver uses the 1d2v solver only for escaping electrons in the loss-cone layer near the body and uses a Fokker–Planck model in position and total energy elsewhere. It conserves particles exactly across the interface and uses 1.6–5.3 times fewer cells. Finally, we discuss the next steps toward parallel, implicit simulations of higher-dimensional problems.

## Highlights

- Separate adaptive meshes for electrons, ions and fields in one code
- Kinetic, reduced and fluid species solvers behind one fixed interface
- Kinetic or fluid model chosen cell by cell for ions and electrons

- Noise-free kinetic model of a capacitive discharge
- Coupled Boltzmann and Fokker–Planck electron solvers around a spherical probe

## 1. Introduction

Weakly ionized plasmas contain neutral particles, ions, and electrons whose masses, velocities, and collision rates differ by orders of magnitude. A simulation tool that accurately resolves each species' transport must therefore adapt the computational mesh in configuration and velocity space and select appropriate kinetic or fluid descriptions based on local criteria. Garcia et al. [1] named this combination Adaptive Mesh and Algorithm Refinement (AMAR) and embedded a Direct Simulation Monte Carlo (DSMC) patch into a continuum solver on an adaptive mesh. In this paper, we describe the plasma extension of the AMAR methodology we have developed over the past two decades and demonstrate its newest element. This element consists of separate adaptive Cartesian meshes for electrons, ions, and the electric field, each with different phase-space dimensionality, refinement criteria, and time steps per solver, all in one executable.

Computational fluid dynamics evolved from body-conforming structured and unstructured meshes to Cartesian meshes with embedded boundaries, making automatic mesh generation around complex shapes practical [2]. For kinetic solvers, Cartesian meshes play a second role, since the discrete velocity method (DVM) represents the distribution function on a Cartesian velocity grid, and the conservative projection method [3,4] evaluates the Boltzmann collision integral on that grid while conserving mass, momentum, and energy exactly. The Unified Flow Solver (UFS) [5] combines an adaptive Cartesian mesh in configuration space, refined automatically around embedded boundaries and solution gradients, with a direct Boltzmann solver and fluid solvers. UFS uses cell-by-cell continuum-breakdown criteria to select the Boltzmann and fluid solvers in each cell. Kinetic patches are introduced and removed dynamically, confining the molecular-scale solution to the regions where it is needed. The UFS was subsequently extended to reacting gas mixtures with chemistry and to the Wang Chang–Uhlenbeck equation for molecular gases [6,7].

We have previously carried the adaptive Cartesian mesh into velocity space with adaptive mesh in phase space (AMPS) [8,9]. A tree-of-trees data structure holds an octree in configuration space whose cells each own an octree in velocity space, generated on the fly and adapted to the local distribution function. Mappings between the velocity trees of neighboring cells make the advection conservative. For symmetric problems, we can treat phase space without splitting configuration and velocity coordinates. We described Boltzmann–Fokker–Planck solvers on a single adaptive Cartesian mesh in two- and three-dimensional phase space in [10,11], using spherical coordinates in velocity space with the Lorentz and Fokker–Planck collision operators. The benefits and limitations of the non-split approach are discussed in Section 2.

The 2012 review [6] outlined the challenges of extending AMAR from gases to plasmas. The scale disparity between species calls for different sensors that select kinetic and fluid models for electrons, ions, and neutral species. Furthermore, depending on electron energy, electron groups may require distinct physical models. However, one requirement has remained unresolved: electrons, heavy particles, and fields should each be computed on their own mesh. Section 2 describes how PAMAR does this.

Several extensions of the UFS framework prepared the ground for the present work. The extension to gas mixtures [7] addressed the mass-disparity challenge. The velocity grids of light and heavy components differ in extent and resolution, and we evaluated the collision integrals on a momentum mesh rather than a velocity mesh. We also added particle-based kinetic solvers to the UFS framework: a DSMC solver that uses a single octree mesh for particle movement, collisions, and statistics collection [12], and an electrostatic particle-in-cell (PIC) solver with an adaptive Cartesian mesh [13]. We implemented a mesoscopic solver based on the lattice Boltzmann method (LBM) alongside the DVM and DSMC solvers [14]. For computational electromagnetics, we developed a discontinuous Galerkin time-domain (DGTD) Maxwell solver on a dynamically adaptive Cartesian mesh [15]. Nonconformal refinement near material interfaces removes the stair-casing error of uniform Cartesian grids without losing efficiency, and the mesh follows propagating wave fronts.

The first plasma solvers in the AMAR framework were fluid models [6]. Later, we developed an implicit, fully coupled fluid-plasma solver with an adaptive Cartesian mesh [16]. It solves the drift–diffusion equations for charged species coupled to Poisson's equation, along with the electron energy equation and plasma chemistry. Its applications to non-equilibrium gas discharges showed where the fluid description fails. Multi-phase flows with free surfaces were also coupled to plasma. We simulated discharges ignited over liquid electrodes with a fluid plasma model and a volume-of-fluid description of the liquid [17], resolving the discharge's multi-scale dynamics and the deformation of the liquid surface. We used a two-stream (forward–backward) electron model to study pulsed breakdown and the modified Paschen curves that result when the voltage rise time is comparable to the electron transit time [18]. The same two-stream formulation, with runaway electrons, is part of the non-split kinetic solvers of [10]. We solve nonlocal electron kinetics in the dynamic regime of AC discharges (where the field period is comparable to the energy relaxation time) using a Fokker–Planck solver in position and total energy, coupled to fluid ions [19,20].

We parallelized the UFS using fine-grained domain decomposition based on space-filling curves, the natural ordering of an octree, which allows us to rebalance the load after each mesh adaptation. Porting an adaptive multi-scale kinetic-fluid code to CPU–GPU systems [14,21,22] raised three difficulties that remain characteristic of the whole approach: the irregular data access of an adaptive Cartesian mesh, the vast difference in computational cost between a kinetic and a fluid cell, and the need to keep all CPUs and GPUs evenly loaded while the grid adapts and the algorithm is refined. We wrote CUDA kernels for three UFS modules, namely, the direct Boltzmann (DVM) solver, the DSMC solver, and the LBM solver. We demonstrated double-digit speedups on a single GPU and good scaling across multiple GPUs with MPI. Section 5 discusses how these lessons carry over to PAMAR, whose solvers currently run sequentially in one process.

PAMAR (Plasma AMAR) is the plasma extension of the methodology above. The central idea is that each plasma species is described by an independent adaptive-mesh solver with its own phase-space dimensionality, refinement criteria, and time step. The species transport is coupled through a field solver in one executable. In the present paper, the physical model for electrons and ions is a build-time switch between a fully kinetic, a reduced-kinetic, and a fluid solver behind one fixed interface. Thus, the model hierarchy (the algorithm-refinement half of AMAR) is a property of the framework, independent of the problem setup. We also demonstrate the ability to select a model dynamically on a cell-by-cell basis, for the ions of the capacitive discharge (Section 3.7) and for the electrons around the sphere (Section 4.6). The framework is built on Basilisk [23,24], a free software system for solving partial differential equations on adaptive Cartesian meshes, targeted to multi-phase flows with free surfaces. Basilisk supplies the quadtree and octree data structures, the wavelet-based adaptation, and the conservative advection schemes that the kinetic solvers use.

PAMAR fills a niche between two established approaches to low-temperature plasma modeling. Particle-in-cell simulations with Monte Carlo collisions (PIC-MCC) [25,26] resolve the full spatio-temporal structure of a discharge. However, they incur statistical noise that scales as the inverse square root of the particle number and is worst in the electron-distribution tail that controls ionization. A recent tutorial case [27] runs about ten billion time steps over a hundred thousand RF cycles for capacitively coupled plasma (CCP) simulations. Fast kinetic models of CCP [28–32] reduce the distribution $f(\boldsymbol{r}, \boldsymbol{v}, t)$ to a function of total energy and run orders of magnitude faster than PIC. However, their assumptions (temporal averaging) fail for the fast electrons whose inelastic collision frequency is comparable with the driving frequency. Eulerian Vlasov solvers [33–35] resolve the tail on a grid, without noise and phase-resolved, but on uniform phase-space meshes their cost grows as the product of the resolutions required by the lightest and the heaviest species. PAMAR resolves the tail on a grid at adaptive cost, each species on its own mesh and with its own physics. Potential applications of PAMAR include streamer and spark discharges and expanding plasmas such as pulsed laser ablation or solar wind.

We present two demonstrations. The first (Section 3) is a one-dimensional transient CCP in argon, computed with two-stream kinetic solvers for electrons and ions on independent adaptive quadtrees in $(x, v)$. The second (Section 4) is a sphere at floating potential in a collisionless or weakly collisional Maxwellian plasma, computed with 1d2v $(r, v, \mu)$ octree solvers for both species. This problem exposes per-species disparities in phase-space requirements, orbital physics, and a boundary whose potential is set by a kinetic flux balance. We compare the mesh-graded solution with an exact method-of-characteristics solution for a steady-state collisionless benchmark. We also confine the 1d2v electron solver to the loss-cone layer near the body and use a Fokker–Planck model elsewhere. Section 2 describes the PAMAR architecture and the equations solved. Section 5 discusses the results and outlines future developments. Section 6 concludes.

## 2. Architecture

### 2.1. Hierarchy of transport models

A hierarchy of transport models describes electrons and ions in a weakly ionized plasma. PAMAR is designed to use these models interchangeably behind one interface. At the top is the kinetic equation for the velocity distribution function $f(\boldsymbol{r}, \boldsymbol{v}, t)$. Because electron and ion collisions with neutrals are linear in $f$, a spherical coordinate system in velocity space, with the speed $v$ (or the kinetic energy $\varepsilon$) and the pitch-angle cosine $\mu$, simplifies the collision operators. Elastic scattering becomes a relaxation of $f$ toward its angular mean (the Lorentz model) or a diffusion in $\mu$ (the Fokker–Planck form of small-angle scattering). Energy loss in quasi-elastic collisions yields a Fokker–Planck operator, while inelastic collisions become shifts along the energy axis at fixed position [10]. For electrons, the angular-moment level lies between the Boltzmann and fluid models. At this level, a few angular moments with appropriate closure produce an accurate description at reduced cost. The two-stream (forward–backward) model of Section 2.3, the two-cone model of Section 2.5, and the Lorentz model are members of this level. Each replaces the distribution's pitch-angle dependence with a few angular moments and retains its full speed dependence, which distinguishes a reduced kinetic model from a fluid closure. Section 5 discusses the importance of angular-moment description for electron kinetics in collisional low-temperature plasma. For low-energy electrons, the Lorentz model is adequate, and the isotropic part of the distribution obeys a Fokker–Planck kinetic equation in the $(\boldsymbol{r}, \varepsilon)$ space. Furthermore, when the energy relaxation length exceeds the plasma size, the isotropic part depends on the total energy $\mathcal{E} = \varepsilon - e\phi$ and obeys a Fokker–Planck (diffusion-like) equation in the $(\boldsymbol{r}, \mathcal{E})$ space [19,31,36]. In low-temperature plasmas, the ion temperature is close to the gas temperature, and charge-exchange collisions often dominate ion transport. Fluid models for ions are usually based on two (drift–diffusion) or three moments that retain inertia [16]. For electrons, two types of fluid models exist: the drift–diffusion model [37,38] and one that includes energy balance [16]. For ions, a similar ladder appears in Section 2.5: kinetic, angular moments, and velocity-moment fluid solvers, with the drift–diffusion model as a limiting case. The adequate level depends on the origin of the ion velocity spread. In the planar discharge, ions are born at rest and acquire their spread along the field through charge exchange, so the two-stream model is the reduced level that captures this (Section 2.3). Around the sphere, the ions arrive from the reservoir with a thermal spread of impact parameters. A cold radial model gives purely radial motion, and charge exchange does not restore the angular momentum. The two-cone model is the reduced level there (Sections 4.4 and 4.5). Because the electron temperature far exceeds the ion temperature, the ions drift faster than their thermal speed over most of the plasma, but the flow becomes supersonic only at the sheath edge, where the drift reaches the ion sound speed (the Bohm criterion, Section 3.7).

Figure 1 summarizes the hierarchy of electron transport models ordered by the characteristic scales that select between them.

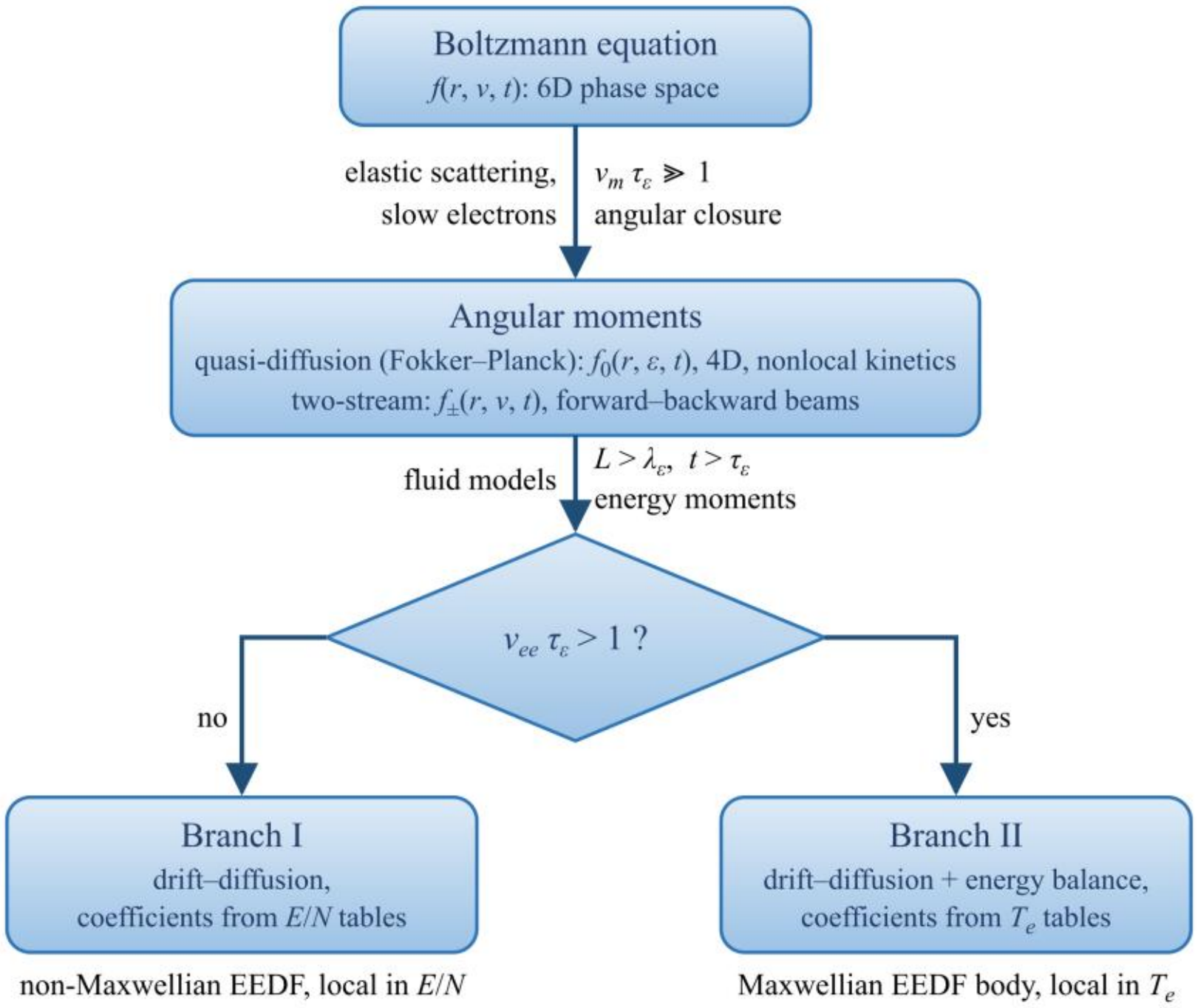

**Figure 1.** Hierarchy of electron transport models by their characteristic scales. An angular closure reduces the Boltzmann equation in the six-dimensional $(\mathbf{r}, \mathbf{v})$ space to the angular-moment level when $\nu_m \tau_\varepsilon \gg 1$ (slow electrons, elastic scattering): the quasi-diffusion (Fokker–Planck) equation for $f_0(\mathbf{r}, \varepsilon, t)$ in the four-dimensional space of position and energy, and the two-stream model for the forward and backward beams $f_\pm(\mathbf{r}, v, t)$ used in Section 3. Energy moments reduce it further to fluid equations when the spatial and temporal scales $L$, $t$ exceed the energy-relaxation length $\lambda_\varepsilon$ and time $\tau_\varepsilon$. Coulomb collisions ($\nu_{ee}\tau_\varepsilon > 1$) Maxwellize the body of the EEDF and justify the fluid model with an electron energy equation, and coefficients tabulated against $T_e$; otherwise, the EEDF and the coefficients are local functions of $E/N$. The species-solver interface of PAMAR is designed so that any member of this hierarchy can stand behind it.

## 2.2. The multi-solver architecture

PAMAR is built on the Basilisk framework [23], extended by a coupling mechanism that lets several independent adaptive-mesh solvers coexist in one executable. Each plasma species has its own solver, with its own phase-space mesh, dimensionality, refinement criteria, and time step. Electrons and ions can each have several species solvers with different physical models (kinetic, angular moments, fluid). A field solver computes the electric field and orchestrates the solver selection and coupling. Technically, each species solver is compiled separately (with the objcopy utility), so the Basilisk runtimes never collide. The field solver sees a species solver only through a fixed interface of six functions: advance to a given time, stop, return the species density and flux at a point, return the species solver's stable time step, and fill the field solver's cell-averaged profiles in one sweep of the species-solver grid. Because the interface is fixed, the species model is a build-time switch behind an unchanged field solver. In particular, the same field solver and electron solver link against a fully kinetic, reduced-kinetic, or fluid ion solver, and the same interface serves 1d1v and 1d2v solvers alike. Couplings other than the electrostatic one used here would require extending the interface. This is the algorithm-refinement half of the AMAR technique [5,6], implemented here at the species level. Section 3.7 demonstrates it at the cell level for ions, with a hybrid solver that is kinetic in the sheaths and fluid in the bulk. Section 4.6 does so for the electrons around the sphere, with a species solver that is kinetic only in the loss-cone layer and a Fokker–Planck model elsewhere. We designed this architecture to select solvers cell by cell for every species, as UFS does for neutral species. The selection is not restricted to one solver per cell. Two solvers for the same species can advance in the whole domain at the same time (one supplying the field solver and the other running alongside it), and the choice between them can be made in time as well as in space (Section 5). The hybrid ion solver of Section 3.7 is built in this way. It carries both descriptions in every cell and chooses per cell which one the field solver reads.

We use two demonstration configurations in this paper. In the planar CCP case, the electron and ion solvers are 1d1v kinetic solvers on independent adaptive quadtrees in $(x, v)$ and the field solver solves the one-dimensional Poisson equation. In the spherical case, the kinetic solvers are 1d2v solvers on $(r, v, \mu)$ octrees, with $\mu$ the pitch-angle cosine, and the field solver solves the radial Poisson equation with a floating or fixed-potential wall. In both cases, the velocity-space resolution requirements of the two species differ by orders of magnitude: the cold-ion distribution occupies $v \leq v_{th,i} = v_{th,e}\sqrt{T_i m/(T_e M)}$, and per-species meshes with per-species velocity normalization (Section 2.7) resolve it at a small fraction of the uniform-mesh cost (Figure 2 and Section 4.1). Table 1 lists the solvers of both configurations and where each is used.

Table 1. The solvers of the two demonstrations: the phase space and mesh of each, its model, and where it is used. Every solver stands behind the same interface of Section 2.2; the hybrids of Sections 3.7 and 4.6 select between two of them cell by cell.

| *Solver* | *Phase space and mesh* | *Model* | *Used in* |
|---|---|---|---|
| *Planar CCP (10 MHz, 50 mTorr, 250 V, 5 cm)* | | | |
| Field | x; adaptive bitree, 512 cells | Poisson, applied RF voltage (Section 2.5) | Section 3 |
| Electrons | (x, v); adaptive quadtree | two-stream kinetic, Eqs. (1)–(3) | 3.1–3.8 |
| Ions, kinetic | (x, v); adaptive quadtree | kinetic with charge exchange, Eq. (4) | 3.4, 3.6, 3.7 |
| Ions, fluid | x; the field cells | two-moment, isothermal, charge-exchange drag, Eq. (6) | 3.6 |
| Ions, hybrid | both, selected per field cell | kinetic where Ti > 1.5 Tg, fluid elsewhere | 3.7, 5 |
| *Sphere in a Maxwellian plasma* | | | |
| Field | r; Laframboise net, 128 shells | radial Poisson, floating or fixed bias, Eqs. (8)–(9) | Section 4 |
| Electrons, kinetic | (r, v, μ); octree, uniform or graded, levels 5–7 | kinetic, elastic scattering, Eq. (7) | 4.2–4.6 |
| Electrons, Fokker–Planck | (r, $\mathscr{E}$); the field shells | Eq. (16); trapped electrons analytic | 4.6 |
| Electrons, hybrid | octree inside ri, Fokker–Planck beyond | selected per cell, Marshak coupling | 4.6 |
| Ions, kinetic | (r, v, μ); adaptive octree | kinetic with charge exchange, Eq. (7) | 4.1–4.5 |
| Ions, fluid | r | radial two-moment | 4.4 |
| Ions, reduced | (r, ε); two-cone (S) | reduced kinetic | 2.5, [48] |
| Exact anchor | orbits | method of characteristics, collisionless | 4.2, 4.6 |

## 2.3. Kinetic equations: two-stream model

For the capacitive discharge, each species is described by a two-stream (forward–backward) distribution $f(x, v, t)$ with the velocity $v \in (-\infty, \infty)$ along the discharge axis. Two-stream models have been used for rarefied-gas dynamics, radiation transport, electron transport in semiconductors (the Shockley model), and fast-electron kinetics in gas discharges. For electrons in discharges, Raizer and Shneider introduced the model as the forward–backward approximation [39,40]. For ions, the model is natural, since the ion drift velocity exceeds the ion thermal velocity in most of the discharge volume. For electrons,

$$\frac{\partial f}{\partial t} + v\frac{\partial f}{\partial x} - \frac{eE}{m}\frac{\partial f}{\partial v} = \beta(\varepsilon)\nu_m(v)[f(-v) - f(v)] + S_{exc} + S_{ion}, \tag{1}$$

where $\nu_m = Nv\sigma_m$ is the elastic collision frequency and $\beta(\varepsilon)$ the backscatter probability per elastic collision. Isotropic scattering corresponds to $\beta = 1/2$ [40]. We use the energy-dependent $\beta(\varepsilon) = (1 - \xi(\varepsilon))/2$ with the screened-Rutherford anisotropy parameter $\xi(\varepsilon)$ of Okhrimovskyy et al. [41], tuned to the argon momentum-transfer cross section. Thus, forward-dominated scattering at high energy transfers fewer electrons between the streams. Equations (1)–(3) follow formally from the kinetic equation in three-dimensional velocity space by substituting a distribution made of two delta functions in direction. The excitation operator is the two-stream shift form [40]

$$S_{exc} = \nu_{ex}(v + v_1)f(v + v_1) - \nu_{ex}(v)f(v), \tag{2}$$

where $v_1 = \sqrt{2\varepsilon_1/m}$. It differs from the usual form in spherical velocity space, where the volume element is $2\pi v^2 dv$. Ionization uses the equal-energy-sharing form

$$S_{ion} = 2\nu_{ion}(2v + v_{ion})f(2v + v_{ion}) - \nu_{ion}(v)f(v), \tag{3}$$

where $v_{ion} = \sqrt{2\varepsilon_i/m}$ and the energies of the primary and secondary electrons after the ionization event are taken equal, with tabulated (or analytic threshold) argon cross sections (Section 3.1); other forms, in which the source term carries the energy loss without creating new electrons, are discussed in [40,42]. Both operators are implemented in number-conserving form on the non-uniform velocity mesh, with the density $n = \int f dv$. In the reference case of Section 3, the electrodes emit secondary electrons with $\gamma = 0.05$ (Section 3.2).

Ions obey the same kinetic equation with $(e/M, +E)$; their collisions with neutrals are dominated by resonant charge exchange, which is described in the BGK form [43,44]

$$S_{CX} = \nu_{CX}[n_i(x)\varphi_M(v) - f_i(v)], \tag{4}$$

where $\nu_{CX} = N\sigma_{CX}\overline{v}$, $\varphi_M$ is the neutral-temperature Maxwellian normalized to $\int \varphi_M dv = 1$ and $\sigma_{CX} = 3 \times 10^{-15}$ cm$^2$ is constant, i.e., the ion mean free path $\lambda_i = 1/(N\sigma_{CX})$ is velocity independent, as in the kinetic theory of the positive column of Dvinin et al. [45]. The operator conserves ion number and relaxes the ion velocity distribution toward the local neutral Maxwellian; on the adaptive $(x, v)$ mesh it reproduces the known square-root dependence of the mean ion velocity on the field [44]. In the code, the rate is evaluated at the local characteristic ion speed

$$\overline{v}(x) = v_{th,i} + |u_i(x)|, \tag{5}$$

with $u_i = \Gamma_i/n_i$, where the thermal floor represents the neutral-target motion in the cold bulk, and the drift term represents the sheath-accelerated population. This per-cell evaluation preserves the constant-mean-free-path scaling and keeps the update local in velocity space. The ionization source deposits the electron-impact ionization rate (computed by the electron solver and gathered by the field solver) on the neutral Maxwellian: $S_{ion,i} = \nu_{ion,e} n_e(x)\varphi_M(v)$.

The fluid ion solver of the planar case solves the two-moment system

$$\begin{aligned} \frac{\partial n_i}{\partial t} + \frac{\partial (n_i u_i)}{\partial x} &= \langle S_{ion} \rangle, \\ M n_i \left( \frac{\partial u_i}{\partial t} + u_i \frac{\partial u_i}{\partial x} \right) &= e n_i E - \frac{\partial p_i}{\partial x} - M \nu_{CX} n_i u_i, \end{aligned} \tag{6}$$

with the ion pressure $p_i = n_i T_g$ at the gas temperature and the same charge-exchange rate (4)–(5) in drag form; neglecting the inertia terms gives the drift–diffusion model. Thus, the comparison of kinetic and fluid ions in Section 3.6 uses identical electrons, fields, and collision rates; only the ion description differs.

## 2.4. Kinetic equations: spherical geometry

For a sphere of radius $R_p$ in a Maxwellian plasma, the species distributions $f_s(r, v, \mu, t)$ obey

$$\frac{\partial f}{\partial t} + v\mu \frac{\partial f}{\partial r} + (1 - \mu^2) \left[ \frac{v}{r} + \frac{a}{v} \right] \frac{\partial f}{\partial \mu} + \mu a \frac{\partial f}{\partial v} = C[f], \tag{7}$$

with $a = q_s E_r/m_s$. We solve these equations for the conservative variable $Y = f v^2 r^2$, so that the finite-volume update conserves particle number exactly on the adaptive octree. The coefficient $(1 - \mu^2)(v/r + a/v)$ carries orbital physics. It vanishes exactly on the turning-point envelope $v^2 r^2 =$ const of the effective potential, and its sign determines whether orbits are focused or defocused by the combination of centrifugal and electrostatic forces [46].

Charge exchange uses the same constant-mean-free-path characteristic rate as (4)–(5), here applied at the particle speed, $\nu_{CX}(v) = (v_{th,i} + |v|)/\lambda_i$, with removal and thermal-Maxwellian re-emission. All ion solvers (kinetic, reduced, fluid) use the same collision operator. The wall at $R_p$ absorbs both species, and a Maxwellian reservoir at $R_{out}$ supplies the unperturbed plasma.

## 2.5. The field solver, the ion models and the exact anchor

The field solver holds the electrostatic problem. In planar geometry, it solves the one-dimensional Poisson equation $dE/dx = 4\pi e(n_i - n_e)$ (Gaussian units) on an adaptive bitree with the multigrid solver of Basilisk [23], driven by the applied voltage $\phi_w = V_{RF}\sin\omega t$. In spherical geometry, it solves

$$\frac{1}{r^2}\frac{d}{dr}\left(r^2\frac{d\phi}{dr}\right) = -\frac{n_i - n_e}{\lambda_D^2} \tag{8}$$

by a Thomas algorithm on the field solver shells, with $\phi(R_p) = \phi_w$ and a Coulomb-tail Robin condition at $R_{out}$. The wall potential floats according to the current balance

$$\frac{d\phi_w}{dt} = GI_{net}(t), \tag{9}$$

or is held at a fixed bias for current–voltage characteristics. Species moments are gathered as conservative finite-volume sums over each species solver's leaf cells.

Three ion solvers are demonstrated. (i) The kinetic solver, Eq. (7) or its planar counterpart, on an adaptive tree is the reference model, resolving orbital motion, charge exchange and trapping. (ii) A reduced-kinetic 1d1v solver, a two-cone (S) model in the spirit of Svettsov's method for spherical radiative transfer [47], carries finite ion temperature at reduced cost. The exact collisionless occupancy at each $(r, \varepsilon)$ is a two-cone band in $\mu$ obtained by inverting the orbital-barrier functions, and charge exchange populates a trapped-ion bath marched in time. This closure physics is the subject of the companion paper [48]. (iii) The fluid solver is the two-moment isothermal model (6) in spherical geometry, with the geometric pressure source and the drag form of the same charge-exchange rate. By construction, it is blind to orbital physics, which makes the kinetic-versus-fluid comparison meaningful.

We use the method of characteristics (MoC) for the collisionless case. In this method, we obtain the species moments for an arbitrary potential profile in closed form (the energy integrals reduce to incomplete gamma functions) and couple them to Poisson's equation via a damped Newton iteration. The MoC solver reproduces the tables of Laframboise [49] and provides independent validation anchors (floating potentials, current–voltage characteristics), exact runtime audits of the kinetic solvers, and warm-start initial conditions that remove the slow ion transient.

## 2.6. Different meshes for different solvers

Each solver generates and adapts its own adaptive Cartesian mesh (ACM). The electron and ion solvers refine on the wavelet estimate of the discretization error of their own distribution function, in their own phase space. The field solver uses a one-dimensional adaptive bitree mesh in configuration space that is independent of the configuration-space projections of the species meshes and can be refined where the field requires it. In the explicit coupling used here, the species solvers are advanced in each field iteration to the field solver's time, their moments are gathered into the field solver's cells by conservative sums over leaf cells, Poisson's equation is solved, and each species solver reads the new field at its own cell centers. Time marching is therefore a sequence of species solver sub-steps between field solves, with each species solver's step limited by its own CFL condition (Section 2.8).

Two mesh strategies deserve separate mention. First, amplitude-based (wavelet) refinement ignores features in the exponentially small tail of a distribution. A feature-aware criterion refines only where a known phase-space surface (e.g., the reflection/absorption separatrix) passes, placing cells where needed at a fraction of the cost. Second, anisotropic refinement is needed where a boundary layer in one coordinate is much thinner than the cell in the others. We obtain it with a graded (mapped) coordinate instead of octree refinement, which halves all three cell sizes at once. The radial axis of each octree is uniform in a grid variable $x$ with $r = r(x)$, and the Jacobian $J = dr/dx$ is carried in the conserved variable $Y = fv^2r^2J$, so that the conservative form is preserved exactly and the time step follows the local cell. The Poisson shells of the field solver use the same net, following Laframboise's computation net [49].

## 2.7. Dimensionless variables and normalization

Each configuration is non-dimensionalized with per-species velocity scales, so that both species' distributions are of order unity on their own meshes. This normalization is part of the per-species mesh strategy.

**Planar CCP.** All physical constants are in Gaussian units, and temperatures are in energy units throughout, so that Boltzmann's constant does not appear. Four parameters define the code units: the length $r_0$ (= 1.25 cm; gap $L_0 = 4r_0$), the time $t_0 = 1/f_{RF}$, and two velocity scales, $v_{e0}$ for electrons, a fixed multiple of the ionization-threshold speed ($v_{e0} = 5\sqrt{2\varepsilon_i/m}$, so that the inelastic thresholds sit at fixed grid positions), and $v_{i0}$ for ions, a fixed multiple of the ion thermal speed ($v_{i0} = 140v_{th,i}$ for the real-mass argon runs). Grid velocities are $y = v/v_{e0}$ and $y = v/v_{i0}$; the electron kinetic energy recovers as $\varepsilon = \frac{1}{2}m(yv_{e0})^2$. Collision operators enter through dimensionless relaxation times $\tau_k = 1/(\nu_k t_0)$. The code potential absorbs the Debye scale: with $\hat{\lambda}_{De} = \lambda_{De}/r_0$ the dimensionless Debye length at the reference density $n_0$ and temperature $T_{e0}$, the physical potential is $e\phi/T_{e0} = \hat{\phi}/\hat{\lambda}_{De}^2$, and the applied voltage is specified in units of $T_{e0}/e$. This makes the Poisson right-hand side of order unity in the quasi-neutral core, however small $\hat{\lambda}_{De}$ is.

**Spherical probe.** The base units are the electron charge and mass, the sphere radius $R_p$ and a reference speed $u_0$, so the code potential is the potential energy $e\phi$ in units of $m_e u_0^2$. Both reservoirs inject $f \propto \exp(-v^2/v_{th}^2)$ ; with the standard choice $v_{th,e} = 0.5u_0$ , this gives $e\phi/T_e = 8\hat{\phi}$ and $v_{th,i} = v_{th,e}\sqrt{T_i m_e/(T_e M)}$. Because the reservoir amplitude rather than the density is normalized to unity, the far-field density is $n_0 = \pi^{3/2}v_{th,e}^3 = 0.696$ in code units, and the one-sided thermal fluxes are $\Gamma_{th,s} = n_0\overline{v}_s/4$ with $\overline{v} = 2v_{th}/\sqrt{\pi}$, the units in which wall currents are audited. The Poisson coefficient defines the input screening parameter $\lambda_D$ through $\varepsilon_0 \equiv \lambda_D^2$. The true electron Debye length is $\lambda_{De} = \lambda_D\sqrt{T_e/n_0} = 0.424\lambda_D$, and the two should not be confused when comparing with Laframboise's parameters. Phase space is mapped affinely per species, $v = V_S y$ with $V_S$ chosen separately for electrons and ions, and $\mu = M_B(z+1) - 1$. The maps have constant Jacobians, so that the conservative form of Eq. (7) is preserved exactly and every normalization constant stays in physical units.

## 2.8. Time stepping

Each species solver returns its own stable step through the interface, and the field solver decides how to combine the species clocks. The electron step is CFL-limited and the smallest. The ion clock can be locked to it, sub-cycled, decoupled, or (for the reduced solver) eliminated. Independent clocks are used only for steady-state problems. Section 4 describes the options and their costs: the ions of the sphere advance with their own CFL step (mass ratios up to 100), at fixed bias and in the floating runs of Sections 4.2 and 4.4, while a floating wall in a large domain required lock step. For transient problems such as the CCP, we retain lock-step coupling, advancing the ion solver over each electron step and solving the field at every step, so that both species advance with the electron step in all CCP runs of Section 3 (Section 5 reports a test with a separate ion step). In the CCP, the time step is set by the electron CFL condition in velocity space (the acceleration $eE/m$ in the sheath) and is about $10^{-4}$ of the RF period, so that a run of 300 periods takes a few million steps. The fluid ion solver saves about a fifth of the CPU time per period (Section 3.6), with the electron solver setting the step in either case.

# 3. Capacitively coupled plasma

The first demonstration is a symmetric planar capacitively coupled discharge in argon. In this configuration, the three operating regimes of AC discharges (high-frequency, dynamic, and quasi-static) appear as the driving frequency varies [19,50], and both PIC results [27] and fast-model theory [28] exist for it. We relate the results to PIC simulations and fast models: the three electron groups and their different time behavior, the two time-averages of the potential that govern electrons and ions, the ion energy distributions, the frequency dependence, and the difference between kinetic and fluid ions.

## 3.1. Cases, cross sections and cost

All runs are for a gap $L = 5$ cm with a sinusoidal voltage applied to the right electrode and the left electrode grounded (no blocking capacitor). The field is obtained from Poisson's equation on the bitree of the field solver, held at 512 cells in these runs, and electrons and ions are each kinetic on their own adaptive $(x, v)$ quadtree unless stated otherwise. The reference case and the benchmark point use the measured argon cross sections (the Phelps set: momentum-transfer elastic cross section, one lumped excitation at 11.5 eV, ionization at 15.76 eV) with the rates $Nv\sigma(\varepsilon)$ evaluated on the electron velocity grid, and the complete inelastic operator with its loss and gain terms. On one core of a shared workstation, the kinetic reference case costs about 6 min per RF period (6400 time steps), the fluid-ion version 5 min, and the benchmark point 6.5 min. We use three sets:

- The reference case: 10 MHz, 50 mTorr, 250 V amplitude, gas temperature 723 K, real argon mass ratio ($M/m = 73000$). At this pressure, the electron mean free path for momentum transfer is 6 cm at 2 eV and 1 cm at 10 eV, comparable to the gap, while the energy relaxation length for elastic collisions exceeds the gap by two orders of magnitude. The ion mean free path $\lambda_i = 0.5$ cm makes the ions collisional. Electrons are therefore nonlocal, and ions are neither in free fall nor in drift–diffusion. A two-moment ion model was used alongside the kinetic solver.
- The frequency scan: 5, 10, 15, 20 and 25 MHz at 20 mTorr, 500 V amplitude, 300 K and $M/m = 65536$, each run until steady (1000–2000 RF periods).
- The benchmark point of Wilczek et al. [27]: argon at 22.5 mTorr, 13.56 MHz, 500 V.

### 3.2. The high-frequency regime: three groups of electrons

The high-frequency regime ($\omega_{pi} < \omega < \omega_{pe}$) is the most common for low-pressure industrial sources and the most demanding to simulate. This is because the RF sheath dynamics must be resolved at MHz frequencies while the time step stays below the electron plasma period and the electron CFL limit. In PAMAR, the step is limited by the electron CFL condition in velocity space, and at 10 MHz the discharge approaches its periodic state over several hundred RF cycles. After 880 periods, the ionization balances 98% of the ion wall loss, and the center density still drifts by 0.3% per 25 periods, which is slow enough for the period-averaged figures below. Figure 2 shows the electron distribution $f(x, v)$ in phase space, together with the characteristics of Eq. (1), at three phases of the RF period of the reference case (left sheath at its maximum, two sheaths equal, and right sheath at its maximum). It is seen that the distribution tail oscillates in time, and the fastest electrons are anisotropic.

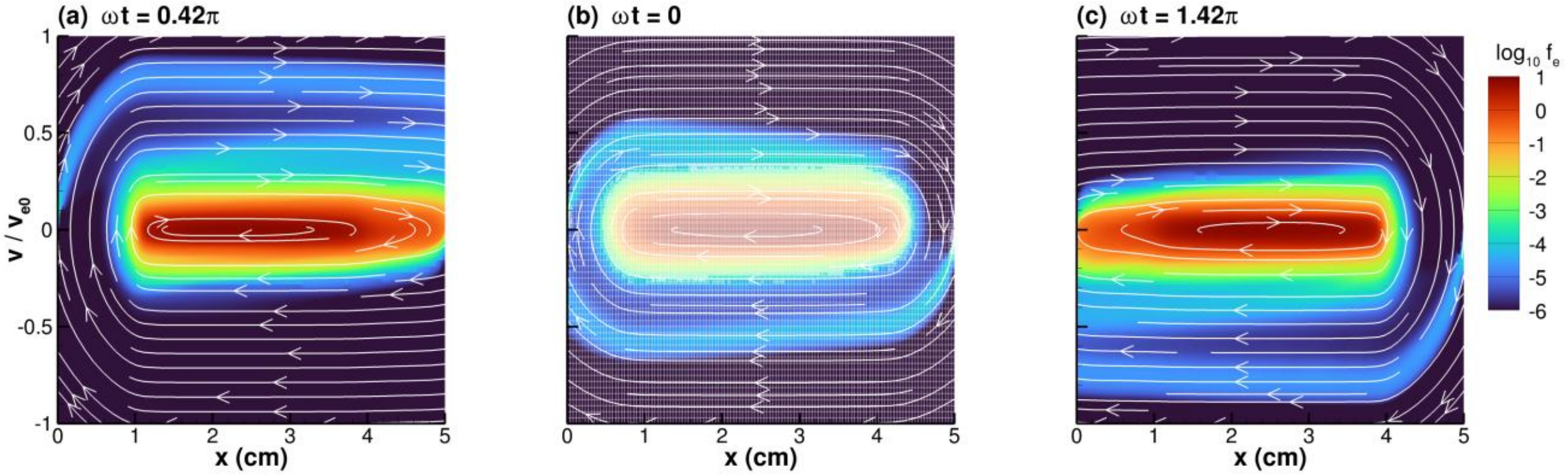

**Figure 2.** Electron distribution $\log_{10} f(x, v)$ (color) and the characteristics of Eq. (1) (white lines) at three phases of the RF period for the reference case (10 MHz, 50 mTorr, 250 V): left sheath maximal ($\omega t = 0.42\pi$), sheaths equal ($\omega t = 0$) and right sheath maximal ($\omega t = 1.42\pi$). Velocity in units of $v_{e0}$ ($\varepsilon = 392.5y^2$ eV). Each panel shows the active domain of the electron quadtree, $0 \le x \le 5$ cm and $-1 \le v/v_{e0} \le 1$ (the 4 × 2 rectangle that the mask cuts from the square root box of the quadtree, see text); the cells are square in the code units and appear with aspect ratio 1:2 in the square panel; the quadtree of the electron solver at that phase is shown in (b) (white cell edges).

Basilisk requires a square domain, so the electron quadtree is set up on the square $0 \le x/r_0 \le 4$, $-1 \le v/v_{e0} \le 3$ ($r_0 = 1.25$ cm), and the cells above $v = v_{e0}$ are masked at start-up, that is, removed from the mesh; $v_{top} = v_{e0}$ is the top of the active domain. The active domain is the $4 \times 2$ rectangle of Figure 2, and the masked half costs neither cells nor time. The range $\pm v_{e0}$, 392.5 eV, is 1.5 times the maximum sheath voltage of the reference case (262 V), and the distribution at the velocity boundary is below $10^{-8}$ of its peak, so the box does not truncate it. All CCP runs use the same box. Higher sheath voltages need a larger $v_{top}$, which is a change of one parameter.

The arms rising from the electrodes in panels (a) and (c) are secondary electrons, emitted with $\gamma = 0.05$ as a 2 eV half-Maxwellian and accelerated by the instantaneous sheath voltage. They cross the bulk at 200–250 eV, at about $10^{-6}$ of the peak of the distribution.

This beam is not resolved. The electron mesh is refined only where $f$ exceeds $10^{-4}$ of its maximum, so the beam, at $10^{-6}$ to $10^{-5}$, sits on the base level of the quadtree, whose cells ($0.031v_{e0}$) are about three times wider than the beam. A feature-aware criterion (Section 2.6) that refines the beam band would resolve it at four to twelve times the present cost per period.

Figure 3 shows the electron density at the three sheath-extremal phases of the final period, its period average, the quasi-stationary ion density, and the ionization rate carried by electrons above the ionization threshold at the same phases. During the ion phase of the sheath, plasma electrons are absent. The plasma–sheath boundary (of the order of the Debye length) oscillates at the applied frequency and is thin compared with the maximum sheath thickness of about 1.4 cm. It is seen that the ionization rate is far from steady. It peaks at about 2.5 times its period average when the sheaths are equal, and the electrons launched by the expanding sheaths cross the bulk, and it falls to about 0.9 times the average at the sheath extrema. Thus, the ionization source oscillates in time and is nonlocal in space.

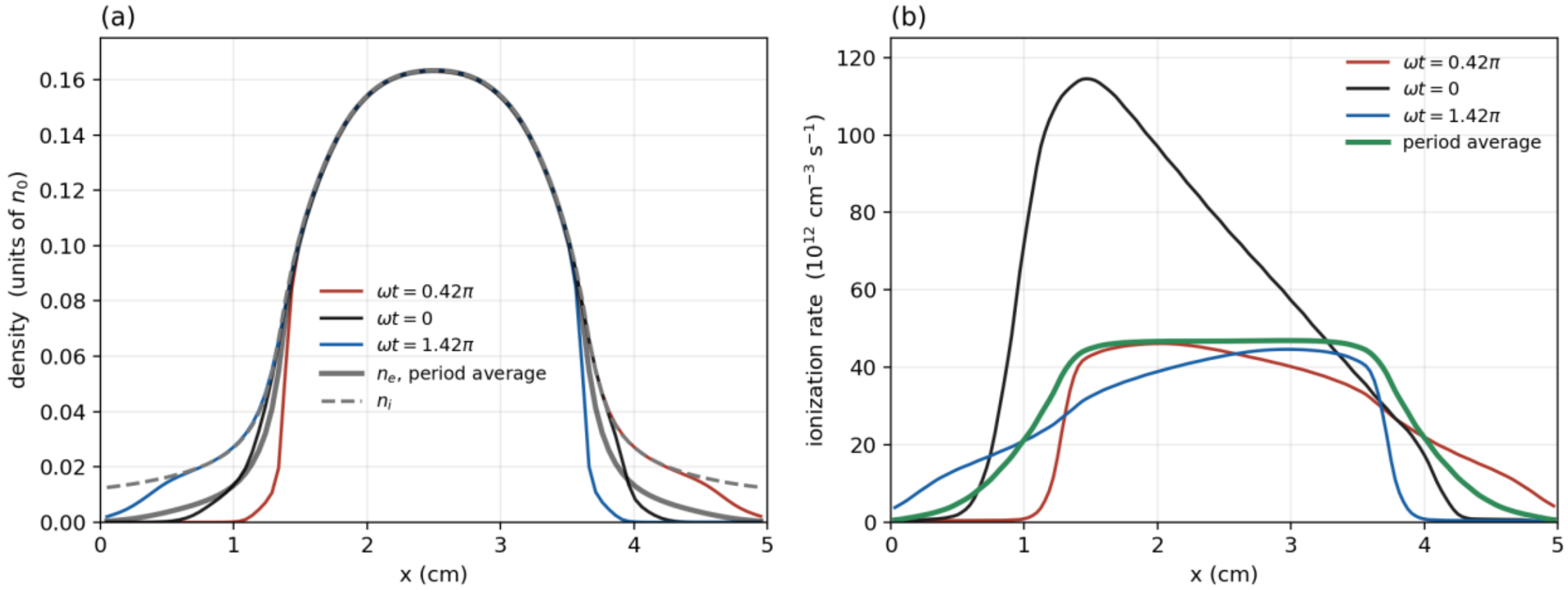

**Figure 3.** (a) Electron density at the three sheath-extremal phases (left sheath maximal, $\omega t = 0.42\pi$; sheaths equal, $\omega t = 0$; right sheath maximal, $\omega t = 1.42\pi$), the period-averaged $n_e$ (grey) and the ion density (dashed). (b) Ionization rate at the same three phases and its period average: the fast-electron population above the ionization threshold peaks when the sheaths are equal. Reference case (10 MHz, 50 mTorr, 250 V)

The velocity distribution at the gap center (Figure 4) shows the three groups of electrons that organize the discharge. Slow electrons ($v/v_{e0} < 0.07$) are trapped in the potential well formed by the time-averaged field. They cannot reach the sheaths, are heated by the small RF field in the plasma, and determine the density profile. Their distribution is almost steady over the period. Intermediate electrons ($0.07 < v/v_{e0} < 0.18$) penetrate the sheaths during the plasma phase and remain trapped by the RF-averaged potential. They are heated by the large sheath field and produce the electron-temperature peaks at the plasma–sheath boundary [51], and their distribution oscillates with the period. Fast electrons ($v/v_{e0} > 0.18$) are strongly anisotropic. They can pass the lowest potential barrier of the period and escape to the electrodes during short intervals near sheath collapse, carry the electron current, and produce the nonlocal, time-modulated ionization. The dashed lines in Figure 4 mark the two boundaries: about 2 eV, where the curvature of the distribution on the logarithmic scale changes sign at the edge of the cold body, and about 13 eV, the lowest barrier between the gap center and an electrode during the period. The same three groups appear in the PIC simulations of [27]. Our simulations resolve a smooth distribution over six decades (Figure 2).

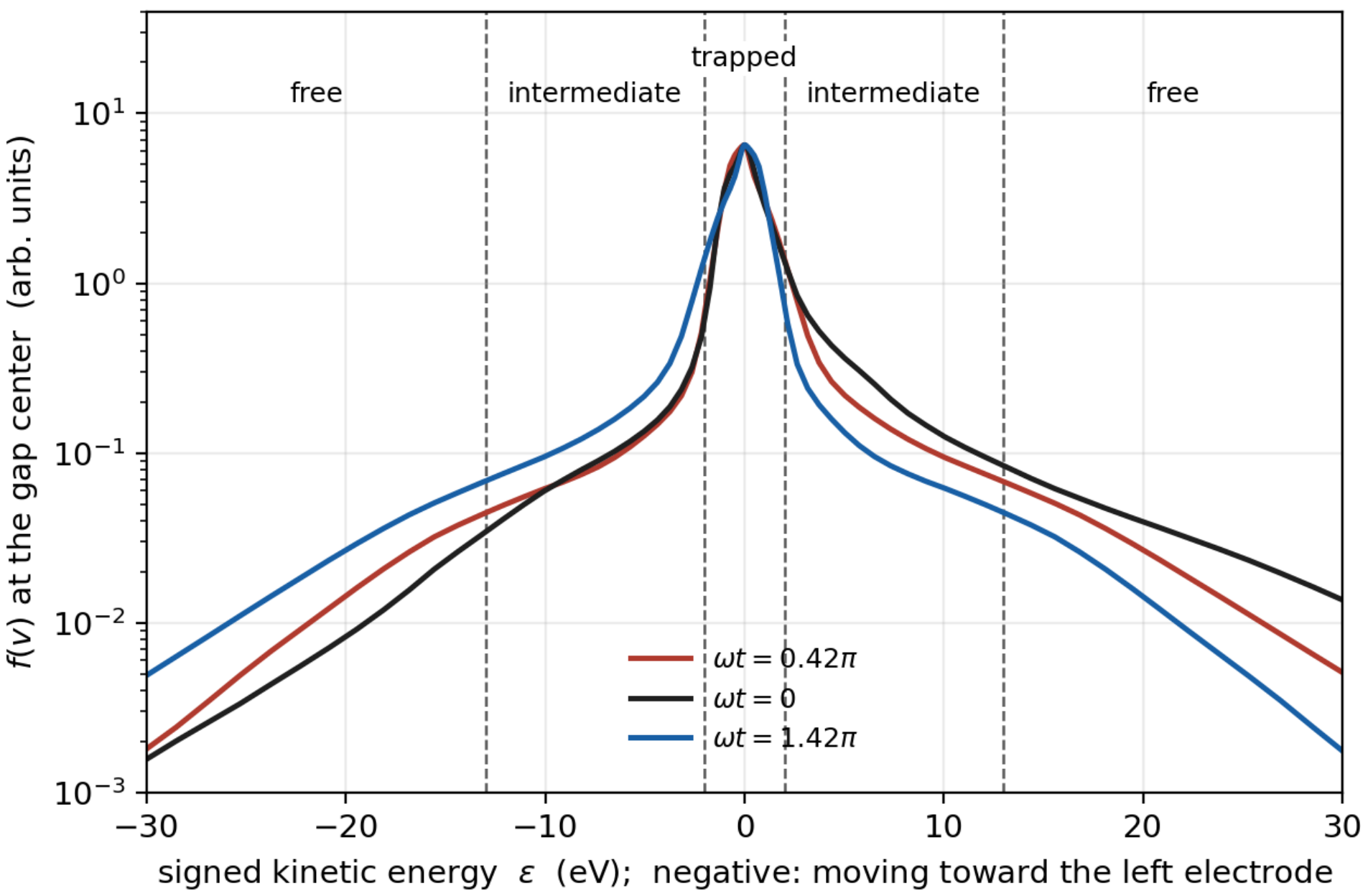

**Figure 4.** Electron velocity distribution at the plasma center at the three phases of Figure 3 (reference case), against the signed kinetic energy (negative for electrons moving toward the left electrode). The body is steady; the tail oscillates and is anisotropic. Dashed lines: approximate boundaries between the trapped, intermediate, and free electrons.

The anisotropic part of the distribution defines the particle currents to the boundaries. The ion current is quasi-steady. The electron current is absent for most of the period and arrives in a brief burst at sheath collapse (near the zero crossing of the total current), so that it never carries the full RF current. The integral balance of electron and ion fluxes over the period is satisfied. The displacement current carries the RF current across the sheath for most of the cycle, and in the quasi-neutral bulk the displacement current is below 1% of the conduction current.

We have observed a high-frequency modulation of $j_D$ near its extrema, which we attribute to the self-excited plasma series resonance of the discharge [52–55]. The ion sheaths act as capacitors, and the quasi-neutral bulk, through electron inertia, acts as an inductor. Together, they form a series resonant circuit with eigenfrequency $\omega_{sr} = \omega_{pe}\sqrt{s/l_b}$ between $\omega_{RF}$ and $\omega_{pe}$. The observed ripple frequency is $8.7 f_{RF} = 87$ MHz, between $f_{RF} = 10$ MHz and $f_{pe} = 257$ MHz at the peak density. The mode is kicked once per cycle by the nonlinear sheath motion, sharpest at sheath collapse, and rings with weak damping at 50 mTorr, where the electron–neutral collision frequency is far below $\omega_{sr}$.

## 3.3. The time averages of the potential

The theory of the capacitive discharge [28] decomposes the electric field into an oscillating part $\tilde{E}(x,t)$ and a slow ambipolar part $\overline{E}(x) = -d\overline{\phi}/dx$. The theory distinguishes two averages of the potential that control the kinetics of electrons and ions. The full-period average contains the ion space charge of the sheaths and gives the deep DC sheath drop that accelerates ions to the electrodes. The average over the plasma phase alone is taken at each position only during the times when that position is quasi-neutral (the interval between the passages of the moving plasma–sheath boundary). This average is the potential that confines the electrons, and it is an order of magnitude shallower. Figure 5a shows both averages for the reference case (referenced to the instantaneous plasma center). It is seen that the full-period average gives a sheath drop of about 120 V, and the plasma-phase average gives an ambipolar drop of about 5 V (about 8 V if the plasma phase is defined by $n_e > n_i/2$). The two coincide in the bulk and separate through the sheath, where the plasma-phase fraction falls from one to zero. In the quasi-neutral core, the current is divergence-free, and the oscillating field is Ohmic, $\tilde{E} \propto 1/n_e$. Figure 5b shows that the amplitude of $\tilde{E}$ (obtained from $-\partial\phi/\partial x$) follows that scaling through the bulk and rises by about a factor of 700 toward the sheath edges. The steeper rise at the very edge is due to the space-charge field.

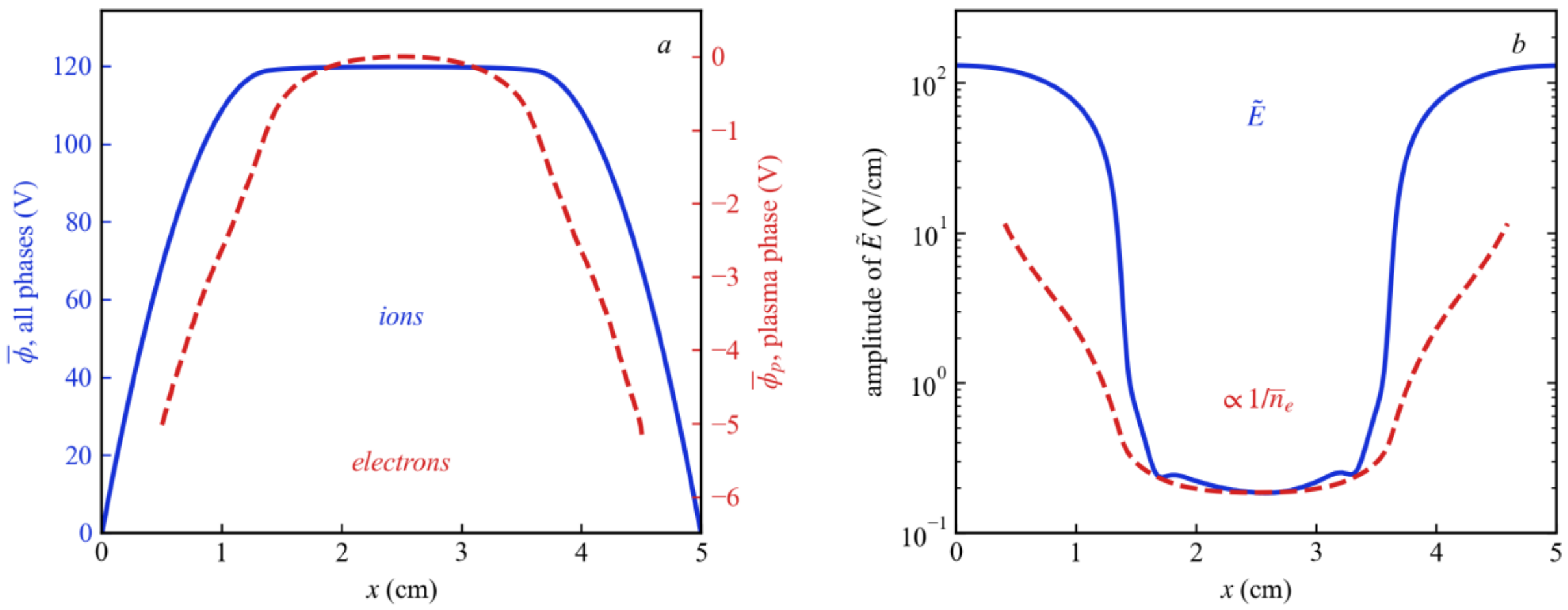

**Figure 5.** (a) The two time-averages of the potential in the reference case (10 MHz, 50 mTorr, 250 V), each RF frame referenced to its instantaneous plasma center. Left axis (blue): the average over all phases, the potential the ions fall through, zero at the electrodes. Right axis (red dashed, own scale): the plasma-phase average that confines the electrons, taken at each position only over the times when the point is quasi-neutral ($|n_e - n_i| < 0.1n_i$). (b) Amplitude of the oscillating (heating) field $\tilde{E}$ from $-\partial\phi/\partial x$ (solid) against the Ohmic scaling $\propto 1/n_e$ (dashed).

## 3.4. Ion transport and the ion energy distribution

Figure 6 shows the ion distribution $f_i(x, v)$ and the ion characteristics at three phases of the RF period. The cold ion bulk lies along $v \approx 0$ across the quasi-neutral core, and ions entering the sheaths accelerate along the S-shaped arms to about $0.4v_{i0}$ at the electrodes. Unlike the electron distribution, the ion pattern is nearly frozen in time, and only the sheath-edge region breathes slightly. The ion energy distribution at the electrode is broad because of charge exchange and ion production inside the sheath. It carries a sequence of secondary peaks produced by ions created by charge exchange at different positions in the oscillating sheath and accelerated by a field modulated during their transit, so their arrival energies bunch. The number of peaks is roughly the number of RF periods an ion takes to cross the sheath [56].

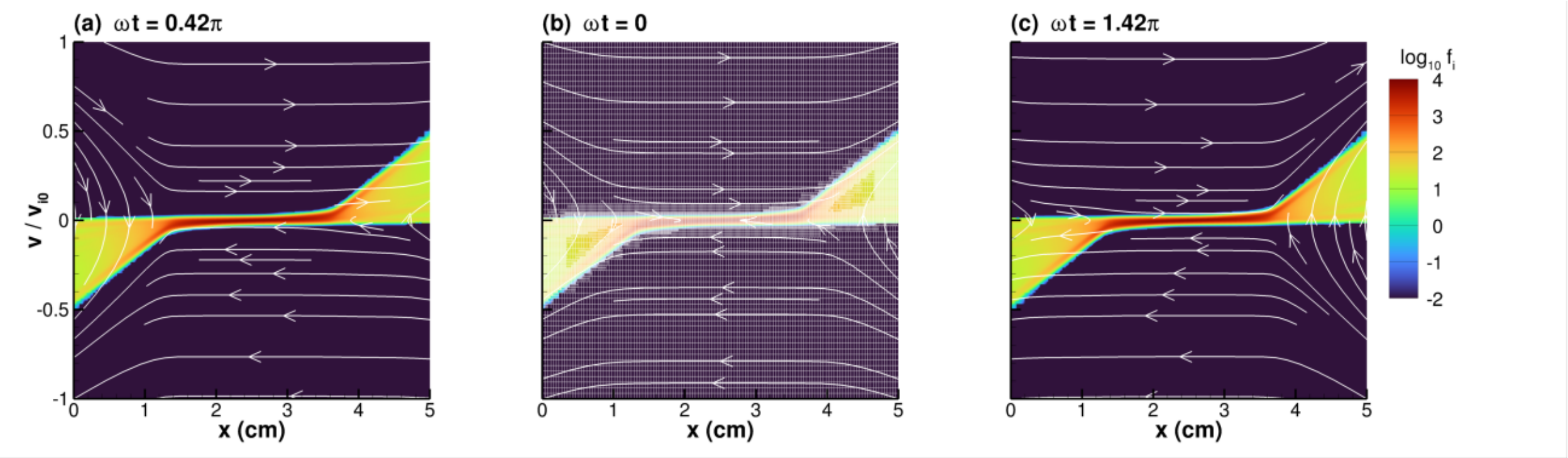

**Figure 6.** Ion distribution $\log_{10} f_i(x, v)$ (color) and the ion characteristics (white lines) at three phases of the RF period of the reference case; velocity in units of $v_{i0}$. Each panel shows the active domain $-1 \leq v/v_{i0} \leq 1$ of the ion solver (the same root box and mask as for the electrons, Section 3.2); the quadtree of the ion solver at that phase is shown in (b) (white cell edges; the cells are square in the code units and appear with aspect ratio 1:2 in the square panel).

Figure 7 shows the ion distribution at the gap center and at the two electrodes for the 10 MHz point of the frequency scan, in the signed-energy representation used for the electrons in Figure 4. Comparing the two figures shows how differently the two species behave. At the center, the ions are a cold Maxwellian at the gas temperature. Their distribution falls to $10^{-4}$ of its peak within 0.3 eV and does not change over the RF period (to 0.2%), whereas the electron distribution at the same point extends to tens of eV, and its tail oscillates with the phase. The ion distribution changes in space instead. At each electrode, it is a one-sided beam at $10^{-3}$ of the center peak that extends to 250 eV, with the main peak at 198 eV, close to the time-averaged sheath voltage of 210 V, and secondary peaks at 20, 67 and 117 eV. The body of the electron distribution, in contrast, depends on total energy and keeps its shape across the plasma (Section 3.8). The two electrodes receive mirror-image ion distributions at all phases. Over the period, the distribution at the electrode varies by up to a factor of four between the peaks (the bands in Figure 7), and the mean energy of the arriving ions varies between 126 and 150 eV around its average of 137 eV, while the ion flux varies by ±9%. The electrons reach each electrode only in a brief burst at sheath collapse (Section 3.2). The ions thus follow the time-averaged field and keep a record of the sheath oscillation in the secondary peaks, while the fast electrons follow the instantaneous field.

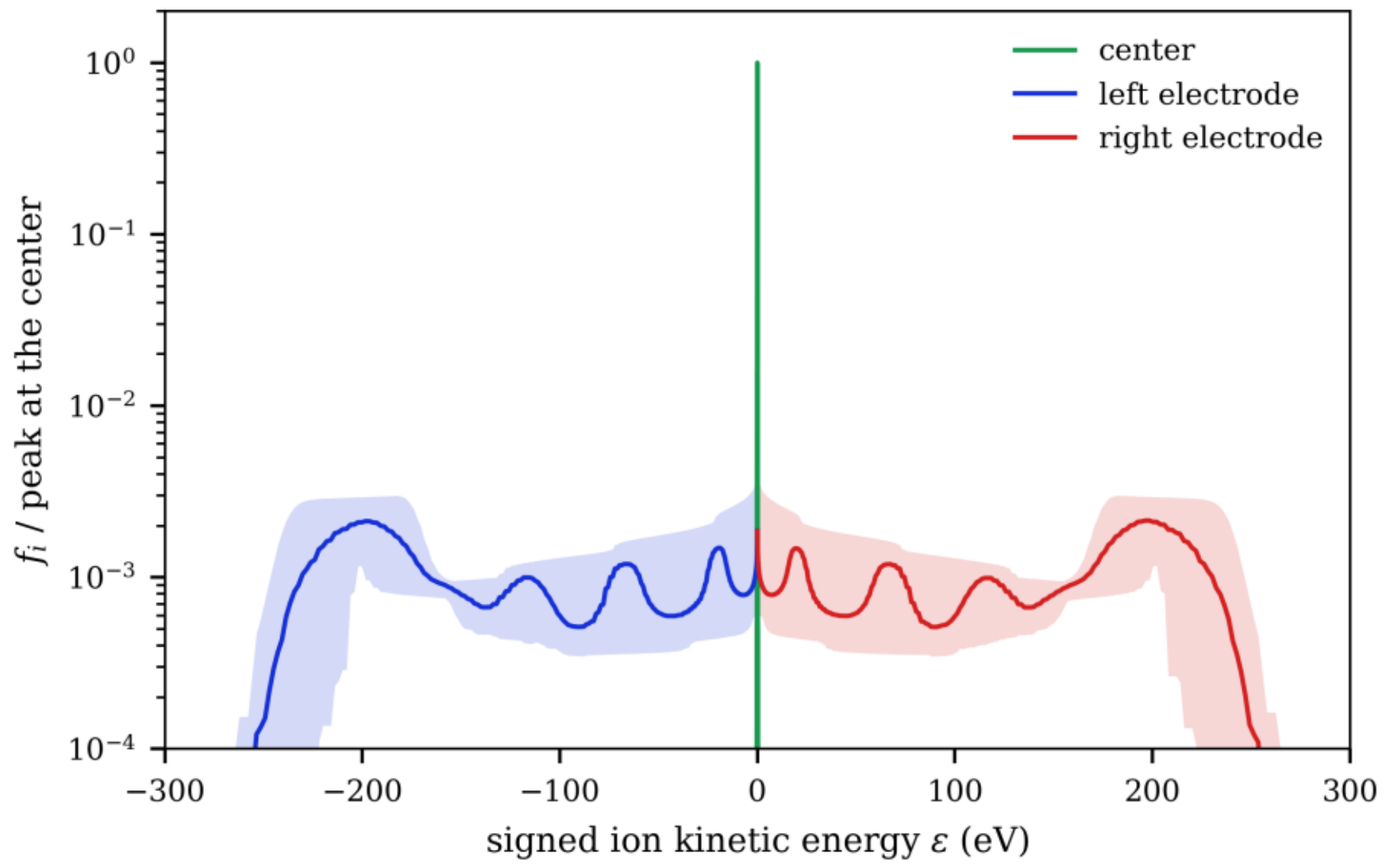

Figure 7. Ion distribution at the gap center (green) and at the left (blue) and right (red) electrodes at 10 MHz (20 mTorr, 500 V, $M/m = 65536$), against the signed kinetic energy (negative for ions moving toward the left electrode), normalized to the peak of the period-averaged distribution at the center. Lines: average over the RF period; bands: range over the period.

## 3.5. The benchmark point: temporal nonlocality

The run at the benchmark point of Wilczek et al. [27] (22.5 mTorr, 13.56 MHz, 500 V, 5 cm, real argon mass, tabulated cross sections) provides the direct, noise-free measurement of the quantity whose neglect defines the validity limit of the fast models. Figure 8 shows the spatio-temporal maps of the slow ($\varepsilon < \varepsilon^*$) and fast ($\varepsilon > \varepsilon_i$) electron densities through the final period with the instantaneous sheath edges. One can see that the slow density is steady and confined between the sheath edges, while the fast density is launched as beams by the expanding sheath and crosses the gap twice per period. Figure 9a shows the period-averaged electron energy probability function at the gap center. It has a two-slope structure, with $T_c = 1.2$ eV for the trapped body and about 5.5 eV above the excitation threshold, where the intermediate group and the tail now share one slope ($T_{int} = 5.5$ eV, $T_h = 5.8$ eV). With the inelastic losses, the tail above the ionization threshold is no longer hotter than the group that feeds it. Figure 9b shows the modulation of $f_0(\varepsilon, t)$ over the period at three energies. The body at 3 eV is modulated by about 25% peak-to-peak and the intermediate group at 8 eV by about 35%, while the tail at 30 eV swings by a factor of about 20 within each half period, as the beams launched by the expanding sheaths pass the center. The plasma-phase-averaged potential gives an electron trap of 1.8 eV in the bulk, and a wall value of 10 eV, against the 220 V DC drop the ions fall through. The cycle-averaged electron temperature is flat at 1.9–2.4 eV inside the trap and rises to about 12–13 eV in the sheaths.

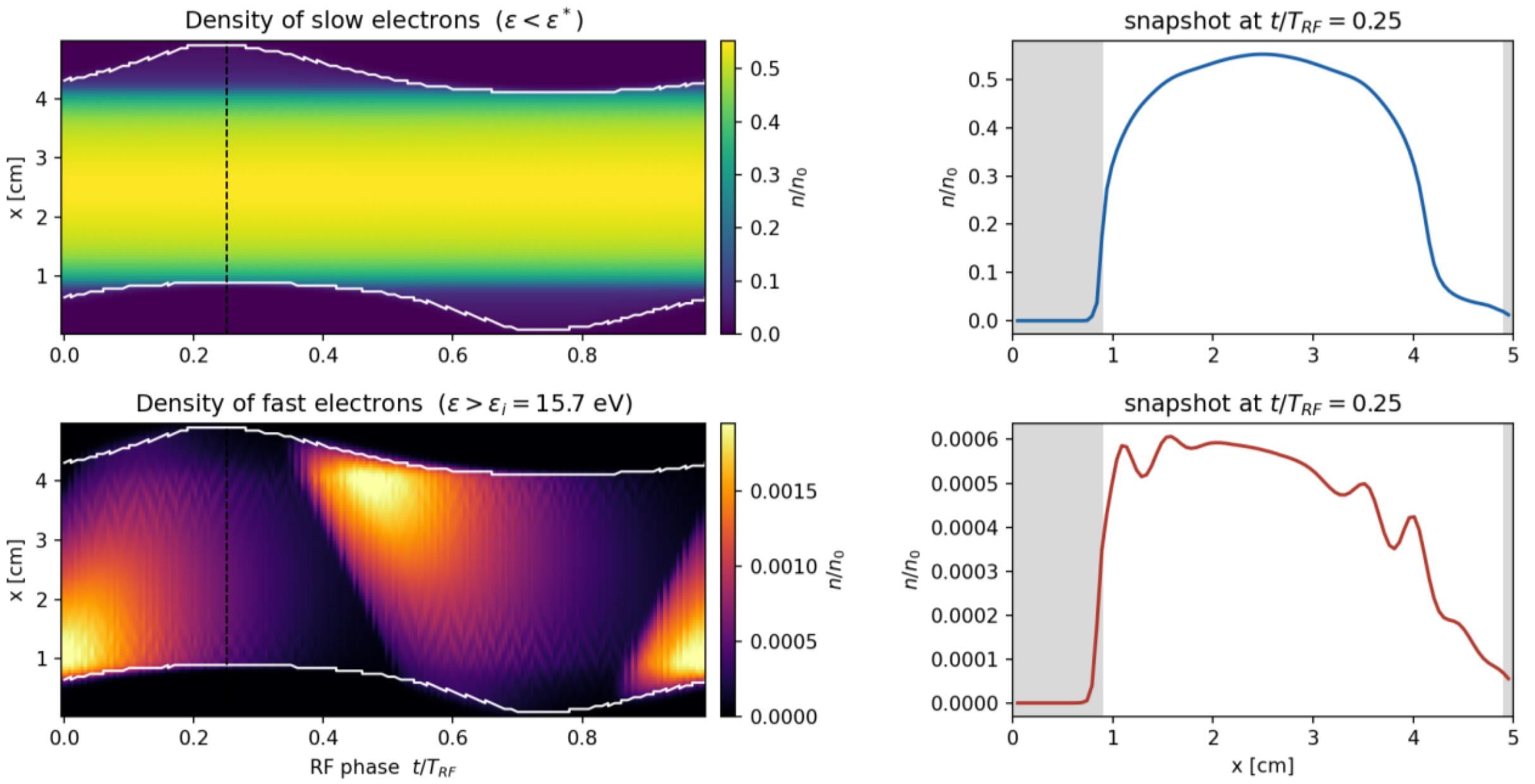

**Figure 8.** Left: density of slow ($\varepsilon < \varepsilon^*$) and fast ($\varepsilon > \varepsilon_i = 15.7$ eV) electrons against position and RF phase, with the instantaneous sheath edges (white). Right: snapshots at $t/T_{RF} = 0.25$ (dashed line on the left); the grey bands are the sheaths. Benchmark point (22.5 mTorr, 13.56 MHz, 500 V).

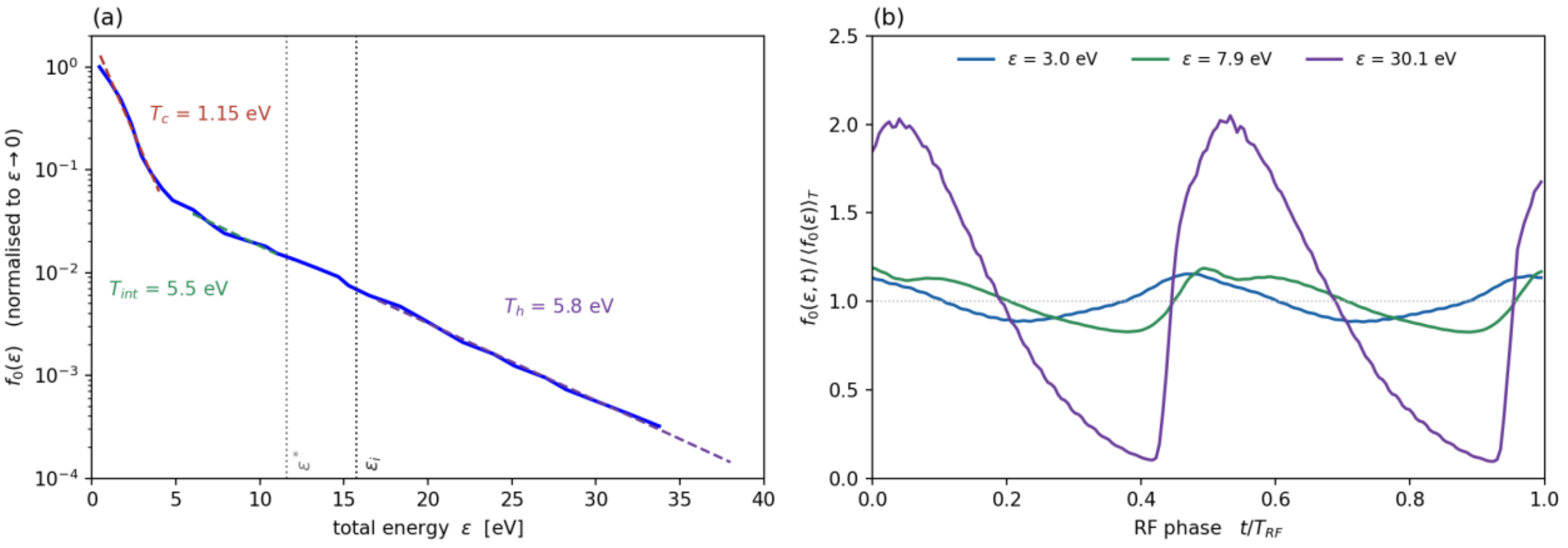

**Figure 9.** (a) Period-averaged electron energy probability function, with the fitted temperatures; dotted lines mark the excitation and ionization thresholds. (b) Temporal modulation of $f_0(\varepsilon, t)$ over the RF period relative to its period average at 3.0, 7.9 and 30.1 eV. The body is modulated weakly and the tail by more than an order of magnitude. (22.5 mTorr, 13.56 MHz, 500 V)

We changed the driving frequency from 5 to 25 MHz at 20 mTorr and 500 V. Between 10 and 25 MHz, the discharge behaves like a conventional high-frequency CCP. The center density and ion flux to the electrodes rise with frequency, the sheaths narrow, and the period-averaged plasma potential (set by the applied voltage) hardly changes. At 5 MHz, the expanding sheaths nearly meet at the gap center, and there is no quasi-neutral bulk. The electron density at the center is modulated at twice the driving frequency, while the ion density (with $\omega_{pi} < \omega$) barely varies. This is the transition from the high-frequency regime to one in which the sheath oscillation spans the gap [19]. The ion energy distributions at the electrodes carry secondary peaks from charge exchange in the oscillating sheath, which shift to lower energies as the frequency rises. A quantitative description of the frequency dependence is beyond the focus of the present paper.

## 3.6. Kinetic versus fluid ions in capacitive discharge

We compare kinetic and fluid ions in the reference case with the ion solver switched at build time, using the same field solver and the same kinetic electrons (Figure 10). We seeded the fluid run with the kinetic ion density and flux at period 590 of the kinetic history and continued both runs for 300 periods. Compared with kinetic ions, the fluid model gives a 23% lower bulk density, and the deficit is spread over the whole bulk (largest at the gap center) rather than concentrated at the sheath edge. The fluid sheath is slightly wider, the period-averaged potentials agree within 2%, and the bulk electron temperature is 10% higher with fluid ions. 10The deficit comes from the isothermal closure, which fails in the sheaths and presheaths where the ions are neither thermal nor a drifting beam. Confined to the fifth of the gap where the ion temperature stays near the gas temperature, the same fluid model costs 1.5% (Section 3.7). The closure itself can be tested on the kinetic solution. The drag term of Eq. (6) with the rate (5), $Mn_i u_i(v_{th,i} + |u_i|)/\lambda_i$, evaluated with the kinetic density and drift, agrees with the momentum lost by the kinetic ions in charge exchange, $(M/\lambda_i)\int v|v|f dv$, to within 4% wherever the drift is below half the ion sound speed $v_s = (T_{e0}/M)^{1/2}$ (although the ion temperature there reaches 4.6 $T_g$). In the sheaths, where the ions are a cold beam plus the slow charge-exchange-born population that a single drifting Maxwellian cannot carry, the drag term falls 20 to 30% short of this momentum loss. What the all-fluid run loses is therefore not the bulk closure but the sheath, fed back on the whole discharge through the sheath width, the electron heating and the ionization. Per RF period, the fluid ions take about a fifth less CPU time than the kinetic ones, because the step is set by the electron solver.

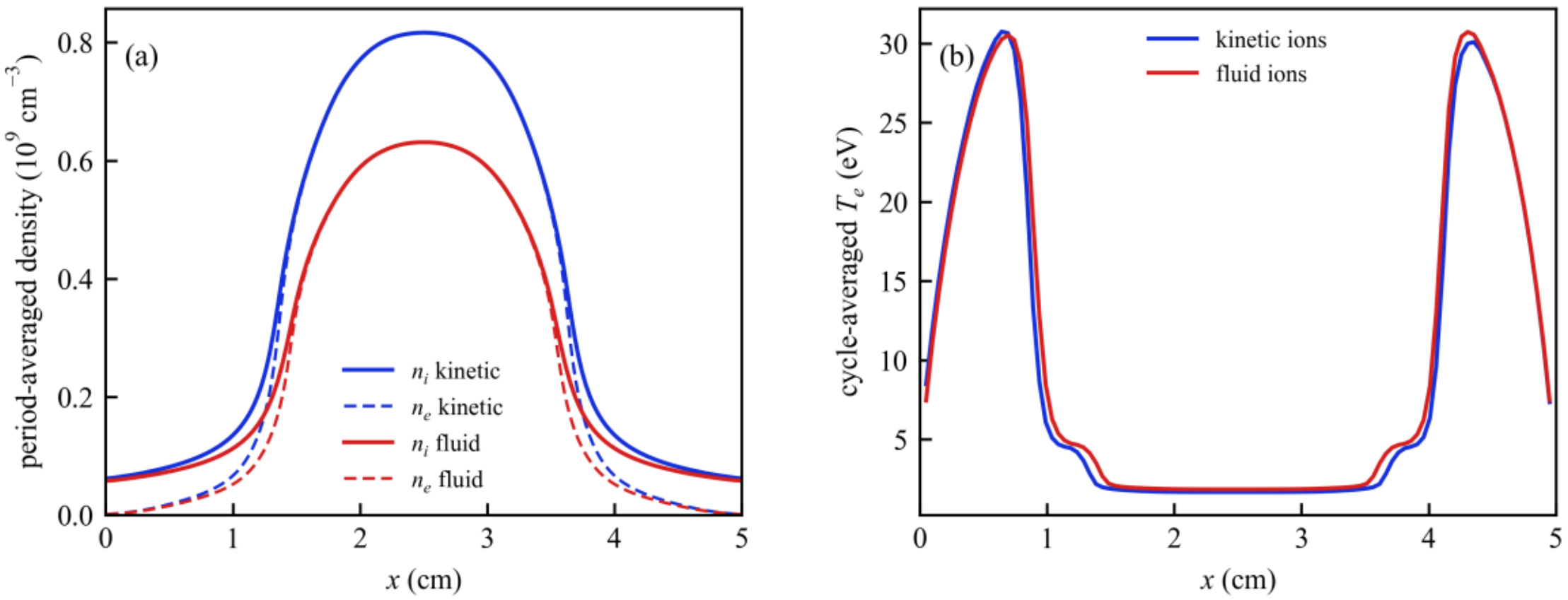

**Figure** 10. Kinetic and fluid ions in the reference case (10 MHz, 50 mTorr, 250 V). (a) Period-averaged densities; (b) cycle-averaged electron temperature.

Because both descriptions are present in every cell of the hybrid solver of Section 3.7, the fluid model can also be run alongside the kinetic ions as a diagnostic: every cell is kinetic, and the fluid model is advanced under the kinetic run's field and ionization source, from the kinetic moments at the restart, without ever supplying the field. This run is a running measurement of the closure error, the drag-law test above extended over 100 periods. Figure 11 shows the relative difference between the two models in density and drift velocity at five positions, sampled once per period. In the sheath, the two descriptions separate within the first ten periods and then stay apart. From period 10 on, the fluid ions are 8 to 10% faster than the kinetic ones (5.41 against 5.03 $v_s$ at the electrodes, at the sampling phase), and this difference does not grow further (Figure 11b). In the bulk, the difference grows throughout the run (Figure 11a). At 1.9 cm, midway between the center and the sheath, the fluid drift exceeds the kinetic one by 15% at period 99, and the fluid density is 6% lower. The sampling phase does not affect these numbers. Within an RF period, the ion velocity in the sheath oscillates in phase in both models, and the period averages give the same ratios. The second moment is fixed by the closure: the fluid ions stay at $T_g$, while the kinetic ion temperature is that of Figure 12b. The closure error thus builds up where the drag-law test located it and accumulates slowly in the bulk, the drift separating the two converged solutions in Figure 10. The coupled solver in the next section builds on this behavior: the fluid description is kept where the twin stays close to the kinetic one.

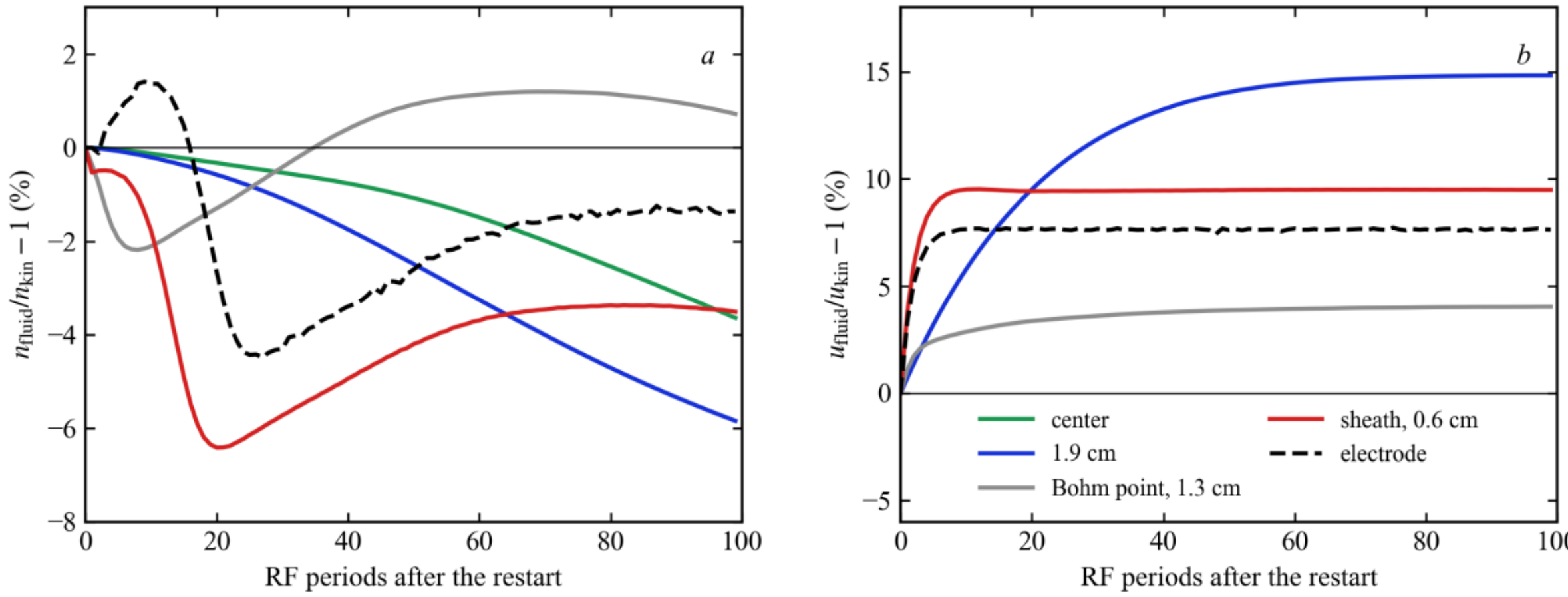

Figure 11. The fluid ion model run alongside the kinetic ions in the reference case (passive twin): relative difference between the fluid and the kinetic (a) ion density and (b) ion drift velocity at the gap center, at 1.9, 1.3 (the Bohm point) and 0.6 cm from the electrode and at the electrode (0.05 cm), against time after the restart (the drift at the center is zero in both models and is not shown), one sample per RF period at $\omega t = 0$. The fluid model is advanced under the kinetic run's field and ionization source without feeding the field.

## 3.7. Per-cell selection of the ion solver

Section 3.6 places the failure of the fluid ion closure at the sheath edge, where the ion distribution is neither thermal nor a drifting beam. This suggests choosing the ion model per cell rather than per run, and applying the algorithm refinement [5,6] at the cell level. We implemented it as a hybrid ion solver behind the same interface. The solver carries the kinetic phase-space tree and, on the field solver's 512 cells, the two-moment fluid ion model of Section 3.6; for each field cell it decides which of the two supplies the ion density and flux to Poisson's equation. Both descriptions are therefore present in every cell, so that the ownership can change from one period to the next. The switching criterion is the ion temperature of the kinetic solution (Figure 12). Once per RF period, the period-averaged ion temperature $T_i = M\langle (v - u_i)^2 \rangle$ of each cell is compared with the gas temperature. Cells with $T_i > 1.5T_g$, plus a buffer of 16 cells (1.6 mm) toward the bulk, are kinetic and the others are fluid, and each kinetic region is kept contiguous from its electrode. The Bohm condition, the natural first choice, places the interface too close to the electrode. At the Bohm point (1.3 cm from each electrode), the kinetic ions are five times hotter than the gas and their distribution is strongly skewed. The Maxwellian at $T_g$ that the fluid feeds into the kinetic region is therefore far too narrow, and the slower ions it supplies raised the hybrid density beside the interface by 20% of the peak and the ion flux to the electrodes by 4% after 100 periods. Only 0.7 cm further in, where the drift has fallen to 0.15 $v_s$, does the ion temperature drop to 1.2 $T_g$ and the isothermal closure of Section 3.6 hold. The temperature criterion finds this point on its own. Starting from an all-kinetic state, the interface settles 2.0 cm from each electrode within two updates and then stays within one cell, so that 21% of the gap is fluid (Figure 12b). Figure 13 compares the hybrid with fully kinetic ions in the reference case, both continued for 100 RF periods from the same late state of the reference run.

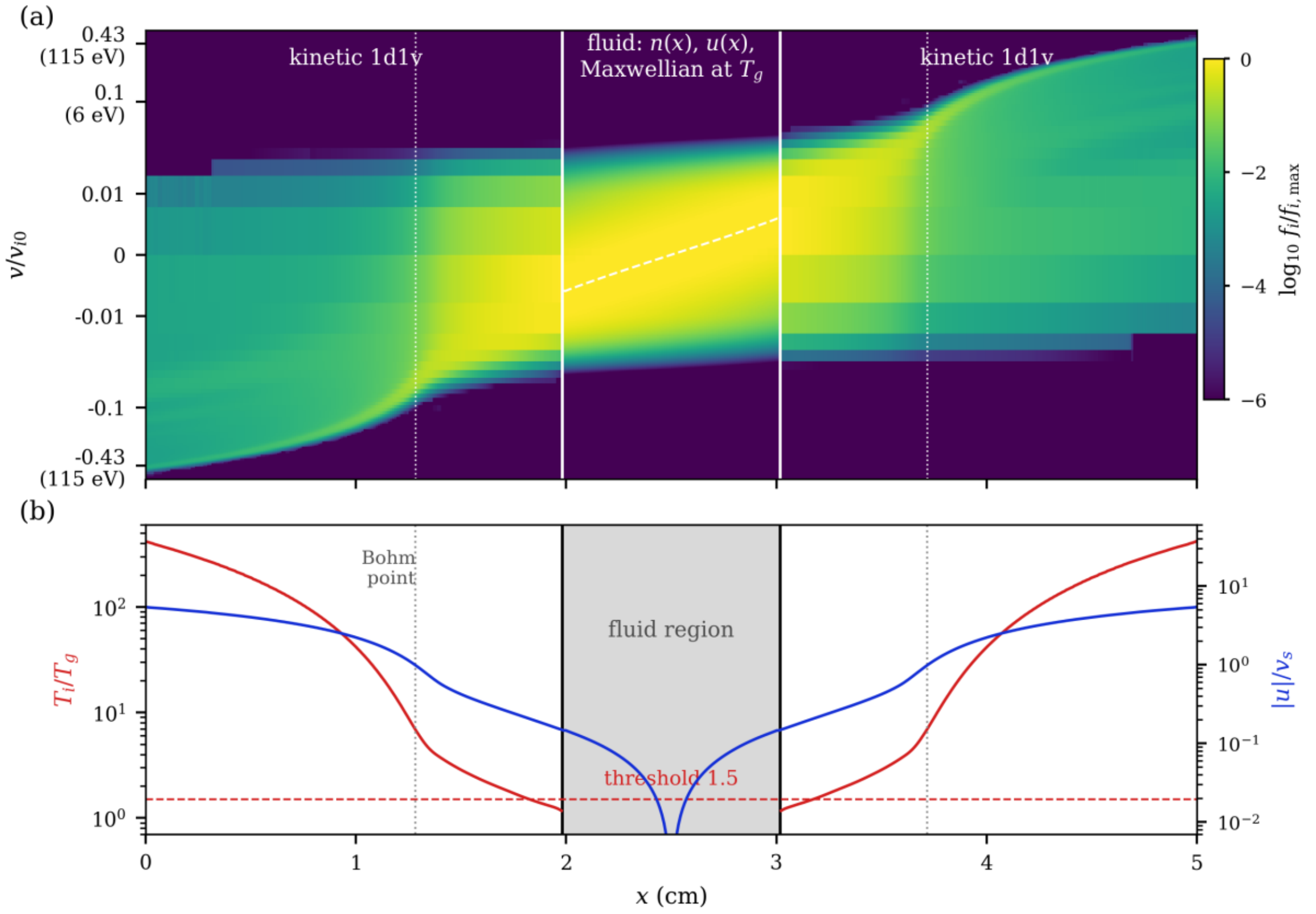

Figure 12. The ion hybrid of the reference case in phase space, 100 RF periods after the restart. (a) The ion distribution $f_i(x, v)$ at the phase of equal sheaths: the kinetic quadtree in the two sheath regions and, between the interfaces (white lines), the drifting Maxwellian at $T_g$ built from the fluid density and drift (dashed: $u(x)$); the velocity axis is logarithmic above $0.01 v_{i0}$ and the energies of the sheath-accelerated ions are marked. (b) The criterion: period-averaged ion temperature of the kinetic solution over the gas temperature (red, kinetic region only) with the 1.5 threshold, and the Mach number $|u|/v_s$ (blue); dotted: the Bohm points; shaded: the fluid region, which begins 16 cells inside the point where the temperature falls below the threshold.

*Interface conditions.* The two descriptions coexist in every cell, which makes the interface conditions explicit. In the fluid cells, the kinetic distribution is replaced at every step by the drifting Maxwellian $f = n_i \varphi_M(v - u_i)$ at the gas temperature. The fluid density and velocity are interpolated to each tree cell, and the Maxwellian is integrated exactly over the cell, so the moments of the overwritten tree equal those of the fluid, regardless of local refinement. Through its interface face, the kinetic region receives the half-range distribution of the ions entering it from the bulk, by the same upwind advection that couples two kinetic cells. The overwritten cells carry no information of their own. Except within eight cells of an interface, their refinement indicator is set to zero, and they coarsen to the base level 7. In the kinetic cells, the fluid state is replaced by the kinetic moments $n_i = \int f dv$, $n_i u_i = \int v f dv$ and the momentum flux $M \int v^2 f dv$, which enter the Rusanov flux at the interface face of the fluid region as its outer state. The kinetic side thus imposes its moments on the fluid, and the fluid imposes the incoming half of the distribution on the kinetic side. Because the drift points toward the electrodes and is subsonic on the fluid side, information crosses the interface mainly from the fluid to the kinetic region. The field solver receives one ion density per cell, so the space charge is single-valued, and the electrons do not see the switch.

*Conservation.* Across the interface, the coupling must conserve the quantities carried by both descriptions: the ion number flux and, for the two-moment fluid, the momentum flux $M n_i u_i^2 + n_i T_i$. The kinetic side passes its full momentum flux, thermal part included, to the fluid. In the other direction, the isothermal fluid has no energy equation and imposes a Maxwellian at $T_g$ on kinetic ions that are 1.2 times hotter. The temperature criterion keeps this mismatch small by construction. Each side still computes its own interface flux (the tree by upwind advection and the fluid by the Rusanov flux), so ion number is conserved only as far as the two agree. Sampled once per period at both interfaces, the ion flux in the last kinetic cell and in the first fluid cell agree to a median of 0.6%, which is the change of the flux across one cell by ionization, so the test resolves no residual. A strictly conservative coupling would use one-sided flux for both sides, as in the refluxing of AMAR [1], and add an energy equation to the fluid.

*Accuracy.* Figure 13 shows that the hybrid closely reproduces the kinetic solution. The period-averaged ion and electron densities agree well everywhere, and the largest difference is no longer a spike at the interface but a 1% deficit spread over the fluid bulk (1.5% at the gap center). The potential agrees to 0.1 V, and the ion flux and the ion energy distributions at the electrodes coincide, with the same peak at 115 eV. The hybrid carries a slightly longer tail beyond 170 eV, four orders of magnitude below the peak. Across the interface, the drift, the temperature, and the heat flux of the hybrid ions coincide with the kinetic ones, so the solution no longer shows where the switch is. The bulk deficit builds up over the first 50 periods and then grows slowly. This is the deficit of the all-fluid run of Section 3.6, confined here to the fifth of the gap that is fluid. Both runs still relax from the restart state (the center density of the kinetic run falls by 1% over the 100 periods), so the comparison is between two trajectories at equal times. The interface does not move with the relaxation.

*Interface position.* Two further tests, each 100 RF periods from the same late state of the reference run as an all-kinetic twin, separate the criterion from the coupling. First, the hybrid solver run with no fluid cells reproduces the all-kinetic run to 0.1% in the density profile and in the ion flux to the electrodes, so the bookkeeping does not perturb the solution. Second, we prescribed the interface at 1.0, 1.3, 1.6, and 2.0 cm from each electrode (60 to 20% of the gap fluid). In this test, the eight fluid cells next to each interface fed the kinetic region not with the Maxwellian at $T_g$ but with the kinetic distribution they already held (rescaled to the fluid density and shifted to the fluid drift). At the Bohm point, this inflow removes the artifacts of the Maxwellian one. The density spike beside the interface falls tenfold, and the excess of the electrode flux disappears. Across the whole sweep, the ion flux to the electrodes stays within 1% of the kinetic twin and the ion energy distributions within the noise of the comparison, so the electrode quantities do not depend on where the switch is. The bulk density, however, depends on it. The deficit at the gap center is 1.5% with the interface at 2.0 cm, where the temperature criterion puts it (the same as with the Maxwellian inflow), and 4 to 5% for every interface at or inside the presheath (still drifting after 100 periods), against 23% for the all-fluid run of Section 3.6. The fluid loses the part of the ionization region where the ion temperature and heat flux are still above their fluid values; the temperature criterion keeps that part kinetic.

*Cost.* There is no saving to report. The fluid columns of the kinetic tree are coarsened rather than removed, and beside the interface they keep their adaptation, so the ion tree is hardly smaller than in the kinetic run. Depending on the kinetic twin used for the comparison (built from the same ion solver or from the newer, faster one of Section 5), the hybrid took between 8% less and 15% more CPU time. The reason is that the hybrid's per-column bookkeeping (scanning the tree, overwriting the fluid columns, and checking their stability) offsets the savings from the fluid cells. The ion solver takes about 40% of the hybrid run time; the electrons dominate, as in Section 3.6. The demonstration therefore shows the mechanism: a solver chosen per cell by the solution itself, at a cost of about 1% in accuracy rather than a speed-up. A speed-up requires removing the kinetic ion tree from the fluid region and applies to problems in which ions dominate the cost, such as multidimensional ion kinetics.

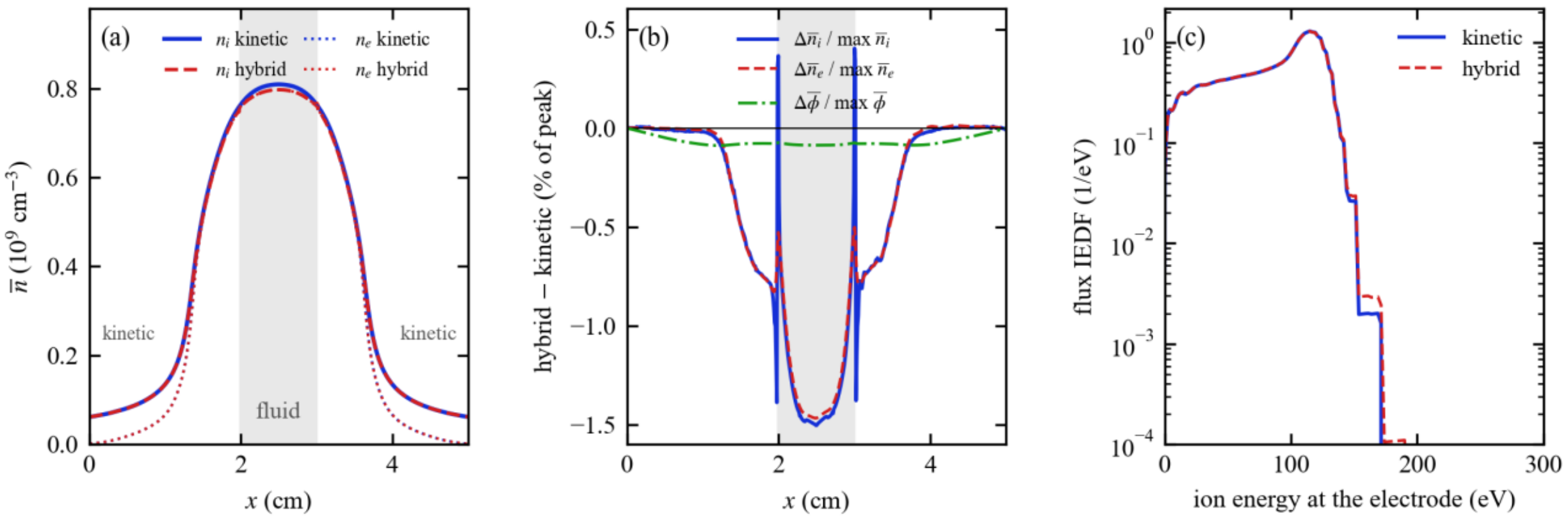

Figure 13. Per-cell selection of the ion solver in the reference case (10 MHz, 50 mTorr, 250 V): hybrid ions (kinetic where the period-averaged ion temperature of the kinetic solution exceeds 1.5 times the gas temperature, fluid elsewhere) against fully kinetic ions, both 100 RF periods after the same restart. (a) Period-averaged ion and electron densities, fluid region shaded; (b) hybrid minus kinetic densities and potential, in percent of their maxima; (c) flux-weighted ion energy distribution at the right electrode.

## 3.8. Fast CCP modeling

The homogenized (fast) description of the capacitive discharge [28,29] rests on three small parameters: (a) $\omega_{pe} \gg max(\sqrt{\nu\omega}, \omega)$, so that the bulk shields the applied voltage and the discharge splits into a quasi-neutral core and ion space-charge sheaths with a moving boundary $x_{1,2}(\omega t)$; (b) $\omega_{pi} \ll \omega$, so that the ion density and the time-averaged ambipolar field are stationary on the RF scale; and (c) $\omega \gg \nu_\varepsilon$, the electron energy-relaxation frequency, so that the body of the electron distribution is time-independent over the period. A fourth parameter, the electron bounce frequency $\Omega \approx v_{th}/L_p \gg \omega$, justifies spatial homogenization, since each electron samples the whole region between its turning points many times per period. Under these conditions, the isotropic part of the distribution depends on the total energy alone, $f_0(x, v, t) \approx F(\mathcal{E})$, and the kinetic equation reduces to an ordinary differential equation in $\mathcal{E}$ with coefficients averaged over the RF period and over the accessible region $[x_-(\mathcal{E}, t), x_+(\mathcal{E}, t)]$,

$$\frac{d}{d\mathcal{E}}\left[\widehat{v\mathcal{D}}_\varepsilon(\mathcal{E})\frac{dF}{d\mathcal{E}} + \widehat{v\mathcal{B}}_\varepsilon(\mathcal{E})F\right] = \sum_k \overline{v\nu_k^*}(\mathcal{E})F(\mathcal{E}) - \sum_k \overline{v\nu_k^*}(\mathcal{E} + \varepsilon_k^*)F(\mathcal{E} + \varepsilon_k^*), \qquad (10)$$

with the energy-diffusion coefficient $\widehat{v\mathcal{D}_\varepsilon} = \left\langle e^2 v^3 \tilde{E}^2 \nu / 3(\omega^2 + \nu^2) \right\rangle$ built from the oscillating field of Figure 5b and the potential well of Figure 5a. Because $\tilde{E}$ rises steeply toward the sheath edges, high-energy electrons reach the high-field periphery, $\widehat{v\mathcal{D}_\varepsilon}$ grows with energy, the tail is depleted, and the two-temperature EEPF of Figure 9a results. The absence of time dependence of $F$ does not contradict the time variation of the electron current to the electrodes, which is carried by the anisotropic part.

The temporal averaging fails where $\omega \leq \nu^*(\varepsilon)$: at low driving frequencies, in molecular gases, at high pressure, and, in every gas, above the inelastic thresholds, where $\nu^*$ rises sharply. There the distribution follows the instantaneous field, the oscillating part $\tilde{f}_0$ is no longer small, and the equation becomes

$$v \frac{\partial F(\mathcal{E}, t)}{\partial t} + \frac{\partial}{\partial \mathcal{E}} \left[ v\mathcal{D}_\varepsilon(\mathcal{E}, t) \frac{\partial F}{\partial \mathcal{E}} + v\mathcal{B}_\varepsilon(\mathcal{E}, t) F \right] = C[F], \tag{11}$$

two-dimensional in $(\mathcal{E}, t)$ rather than three-dimensional in $(x, v, t)$ if the spatial averaging still holds, with the coefficients carrying the instantaneous field. Figures 4 and 9b show directly where this happens in an argon CCP at 10 and 13.56 MHz. The modulation is 10–15% at the excitation threshold, about 50% at the ionization threshold and grows roughly linearly with energy beyond it, so that the ionizing population is temporally nonlocal while the body that sets the density is not. A homogenized model calibrated on the kinetic solution, with $F(\mathcal{E})$ for the body and Eq. (11) for the tail, is the intermediate level of the electron model hierarchy that the PAMAR architecture is designed to accommodate (Section 5), and the kinetic solution identifies the electrons that need it.

## 4. Kinetics of spherical plasma

The second demonstration is the classic kinetic benchmark of an absorbing sphere at near-floating potential in a collisionless or weakly collisional Maxwellian plasma [49,57]. This problem underlies electrostatic probes, dust grains, and spacecraft charging in the ionosphere. It exhibits extreme per-species disparity in phase-space requirements, orbital (angular-momentum) physics, and a floating boundary whose potential is set by a kinetic flux balance. In some respects, it is simpler than the CCP because only the steady state matters, there is no ionization or excitation, and the ion and electron fluxes are conserved. We retain only charge-exchange collisions for ions and elastic collisions of electrons with atoms. We compute the potential self-consistently from Poisson's equation, Eqs. (8)–(9), in Sections 4.2, 4.4 and 4.5 and in the last two rows of Table 2. The mesh and flux tests of Section 4.3 and the other hybrid runs of Section 4.6 use a prescribed Coulomb potential.

**Phase-space demands of the problem.** Two large ratios set the difficulty. The mass ratio and the temperature ratio combine into the velocity-scale separation $v_{th,e}/v_{th,i} = \sqrt{MT_e/(mT_i)}$. For argon ions at room temperature in a $T_e \simeq 2$ eV plasma ($M/m = 73000$, $T_i/T_e \approx 0.01$), the two velocity scales differ by a factor of about 2700. A single uniform phase-space mesh serving both species would need that factor in each velocity dimension. Its time step would remain chained to the electron CFL condition, while the physics of interest (trapped-ion build-up by charge exchange) evolves on the still slower collisional scale $\lambda_i/v_{th,i}$. PAMAR addresses this regime with four elements: (i) per-species meshes with per-species velocity normalization (Section 2.7), (ii) adaptive refinement where the solution structure requires it, (iii) a model ladder that trades the ion velocity grid for a reduced description where that description is exact or calibrated, and (iv) runtime audits of the result. We have chosen the demonstration case ($R_{out} = 3R_p$, $R_p/\lambda_{De} = 7.9$, $M/m = 100$, $T_i = T_e$) so that a fully kinetic reference calculation remains tractable. The reduced model of Section 2.5 extends beyond it [48].

**Two clocks for one steady state.** Laframboise [49] shows by dimensional analysis that the mass ratio enters the computational scheme only when the net current is calculated. Each species' current is a function of $\chi_p = -e\phi_p/T_e$, $R_p/\lambda_D$ and its own temperature ratio alone; the attracted and repelled roles are interchangeable, and $M/m$ re-enters only through the floating condition

$$i_- = i_+ \sqrt{\frac{mT_i}{MT_e}}. \tag{12}$$

Al'pert et al. [57] reach the same point from the solution itself. The steady collisionless distribution and every density that follows from it (hence Poisson's equation) depend on $U(r)/T$ only, and the mass appears in the fluxes. Our MoC solver is built on this structure. It takes $(\chi_p, R_p/\lambda_{De}, T_e/T_i)$, and $M/m$ enters only in the current-balance root find that locates the floating point. For PAMAR, the steady state does not depend on the path by which it is reached, so the two species need not march on a common clock. In lock-step marching, the number of steps to an ion-relaxed state is $\tau_i/dt_e \propto \sqrt{MT_e/mT_i}$ (the electron CFL step applied to ion transits). With independent clocks, the ion solver takes its own CFL step per field iteration, the electron solver stays relaxed against the slowly varying $\phi$ and $n_i$, and the wall floats on the ion clock. The cost then falls by $dt_i/dt_e$, about 2.5 at $M/m = 10$, about 7 at $M/m = 100$ and, extrapolated as $\sqrt{M/m}$, about 200 for argon (the two-clock runs reported here go up to a mass ratio of 100). The same argument admits an artificial mass ratio (this section's $M/m = 3$ to 100). At fixed bias, the profiles are physical, and an artificial mass ratio is obtained by holding the bias at the MoC floating value for the real ratio.

We have verified this independence directly in the kinetic solver. The same prescribed bias $\chi_p = 2$ at $\lambda_{De} = 10R_p$ and $T_e/T_i = 3$, run at $M/m = 3$ and at $M/m = 100$, gives electron densities and potential profiles that agree within 1.5% at every radius and conserved currents that agree within 3%. The residual is attributable to the mesh and the relaxation state, not the mass ratio, since refining the electron mesh at fixed $M/m = 3$ moves the ion current by 9%, several times more than changing the mass ratio by a factor of 33 does. At the fixed-bias benchmark ($\chi_p = 2$, $\lambda_{De} = 10R_p$, $T_e/T_i = 3$, $M/m = 100$), the two-clock run met the steady-state gate after 6.9 ion transits in 43 minutes on one core. The lock-step run, identical in every other respect, was still short of the gate after 6.2 transits and 11.0 hours. The two final states agree to 1% in ion density and are better in electron density and potential. The acceleration is 17.1 times in ion time per wall-second with no measurable change in the converged state, close to the ratio of the two CFL steps, $\sqrt{MT_e/mT_i} = 17.3$. By the same scaling, the argon mass ratio would give about 460 at this temperature ratio; we have not run it. The acceleration is unconditional only at fixed bias. The floating runs of Sections 4.2 and 4.4 ($R_{out} \leq 4.5R_p$) also advance the ions on their own clock and converge to the current balance. However, with a floating wall on $R_{out} = 12R_p$, the free-running ion clock drove a growing oscillation of the bulk potential after about two ion transits, which a lock-step run from the same state damped. For this reason, the floating runs of [48] advance both species in lock step.

**A protocol for steady problems, and its limits.** These results suggest the following protocol: (i) prescribe the bias rather than let the wall float, since at fixed $\chi_p$ the profiles are the physical ones, mesh effects are visible directly instead of being absorbed by a feedback variable, and the run cannot chase a moving target; (ii) use whatever mass ratio is affordable, since at fixed bias the answer does not depend on it; (iii) give each species its own clock, which sacrifices the (now fictitious) transient and recovers the factor $\sqrt{MT_e/mT_i}$; (iv) recover the floating state afterwards, at the physical mass ratio, from Eq. (12) evaluated on the computed characteristic. This protocol has limits. It covers the steady state only, and nothing time-dependent may be read from such a run. The path can be forgotten only where the steady state is unique; with trapped orbits, it is not. With charge exchange (Section 4.4), the trapped population is set by the physical collisional clock ($\lambda_i/v_{th,i}$), which the ion clock must honor, while the electrons can still be relaxed to it.

## 4.1. Per-species meshes

Even at the reduced ratios of the demonstration, the two species see incommensurate phase spaces. The electron distribution spans $v \leq 4v_{th,e}$, while the cold-ion population lives at $v \leq v_{th,i} = v_{th,e}/\sqrt{M/m}$, an order of magnitude smaller. On a common uniform octree, resolving the ions at about 13 cells per $v_{th,i}$ would cost $2.1 \times 10^6$ cells. The adaptive ion octree achieves that resolution with 10.7% of the uniform-mesh cell count (refining onto the cold bulk and the absorption boundary), while the electron solver keeps its own uniform mesh and its own velocity normalization. The field-free solution reproduces the Al'pert density profile $n/n_\infty = \frac{1}{2}[1 + \sqrt{1 - R_p^2/r^2}]$ to better than 2% and the unit current to 95% (uniform level 5) and 97% (graded level 5, Section 4.3); the same gate run by the reduced ion solver (Section 2.5) reproduces it to $5 \times 10^{-5}$ with exact unit flux.

## 4.2. The method of characteristics

For the collisionless problem, Liouville's theorem makes the exact solution computable for any potential profile, since along each orbit the distribution keeps its boundary value. Orbits are classified by the two functions

$$B_{out}(r, \ell^2) = max_{s \geq r}\left[q\phi(s) + M\frac{\ell^2}{2s^2}\right],$$
$$B_{in}(r, \ell^2) = max_{R_p \leq s \leq r}\left[q\phi(s) + M\frac{\ell^2}{2s^2}\right], \qquad (13)$$

and the angular reductions leave one-dimensional energy integrals that evaluate in closed form (incomplete gamma functions). Inbound points are populated if the energy exceeds $B_{out}$, outbound points if it lies between $B_{out}$ and $B_{in}$, the orbit having been turned by the centrifugal barrier before reaching the surface; everything else (the loss cone and the bound orbits) is empty. We have implemented this in our MoC solver with a damped-Newton coupling to Poisson's equation. The solver reproduces the Laframboise tables [49] and the recent independent calculations of Filippov [58]. One species evaluation takes a fraction of a second.

14 shows the converged PAMAR kinetic floating point at the Laframboise-regime screening length (level-6 electrons on the graded mesh, 128 Poisson shells on the same net) against the MoC. Both currents are read from the conservative face fluxes, which are constant across the domain to 0.5% for the ions. The densities agree with the MoC profiles to 0.5% on average, and the potential to 2% of the wall value. The residual sits in the presheath, where the mapped outer cells are coarse.

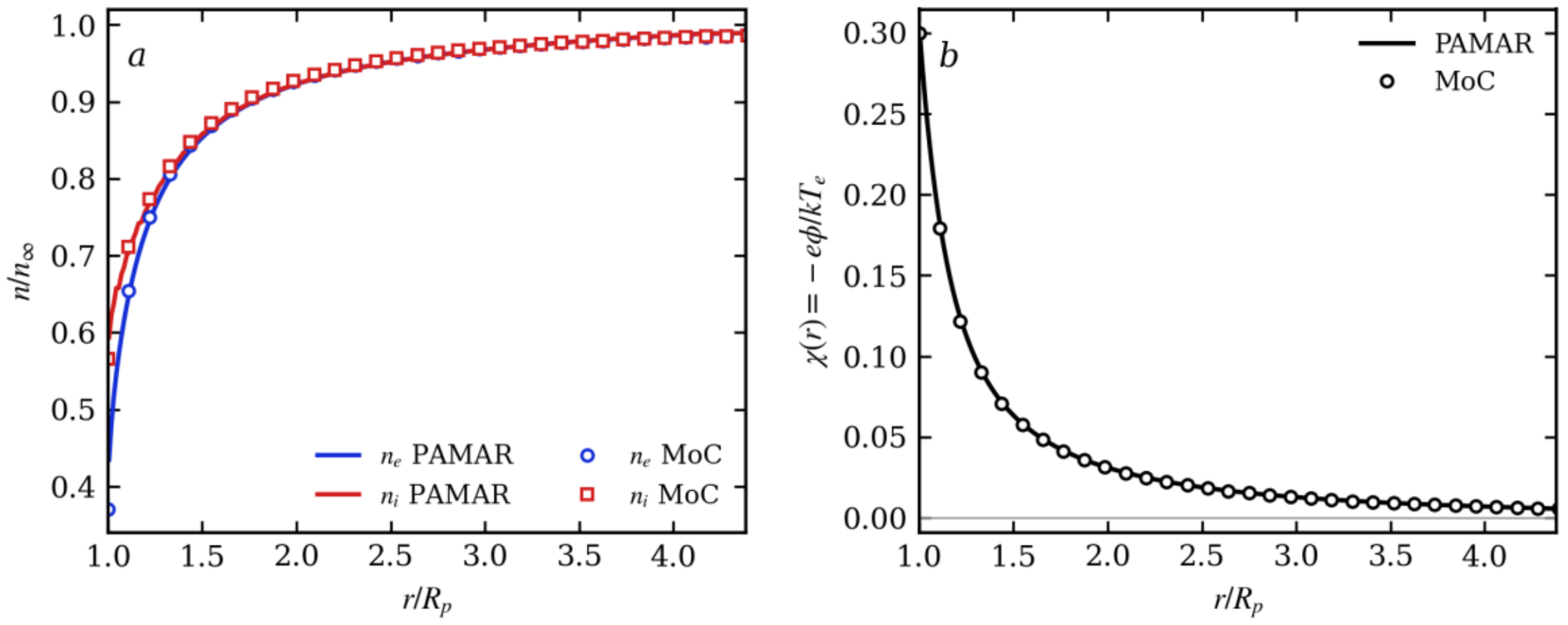

**Figure** 14. The converged kinetic floating point compared with the exact MoC solution ($R_p/\lambda_{De} = 7.9$, $M/m = 3$, $T_i = T_e$, $R_{out} = 4.5R_p$ with the $r^{-2}$ tail, level-6 electrons on the graded mesh, 128 Poisson shells): (a) densities, PAMAR (lines) and MoC (symbols); (b) the potential $\chi(r) = -e\phi/T_e$. Wall: $\chi_f = 0.306$ against 0.300. Currents from the conservative face fluxes: ions 1.276 against 1.283 $\Gamma_{th,i}$ ($-0.5\%$), electrons 0.732 against 0.741 $\Gamma_{th,e}$ ($-1.2\%$).

## 4.3. The electron flux, mesh grading and the wall cell

For a collisionless repelled Maxwellian species, Liouville's theorem fixes two things exactly: the absorbed flux cannot exceed

$$\Gamma_e \leq \Gamma_{th,e} e^{-\chi}, \tag{14}$$

and the total flux $r^2\Gamma(r)$ must be conserved. We check both at runtime with a face-flux audit, which evaluates the radial current through the faces of the phase-space octree from the same face fluxes the transport uses and splits it into its inward and outward parts. In a steady state, the net current must be the same on every face unless particles leave through a velocity boundary. Applied to the electron solver in a prescribed Coulomb field ($\chi_p = 2$, $\lambda_{De} = 10R_p$, $R_{out} = 10R_p$, started from the exact Liouville state), the audit separated three errors. Two of them were boundary conditions, and only the third is resolution.

*(i) Boundary fluxes.* The second-order upwind reconstruction evaluates a boundary face from its ghost cell, and at two faces the ghost was wrong. Both faces now carry exact fluxes, since the wall emits nothing and the reservoir injects the Liouville state from infinity, $v\mu f_\infty(\mathcal{E})$ with $\mathcal{E} = mv^2/2 + q\phi(R_{out})$, through the $r^{-2}$ tail of the field solver. The second boundary error was a leak, in which decelerated electrons left phase space through the $v = 0$ face at their turning points instead of turning through $\mu$. A no-flux face closes the leak. With the three fixes, the net current is constant across all faces to $2 \times 10^{-4}$.

*(ii) The wall cell and mesh grading.* With exact boundary fluxes, the inward flux matches the exact Boltzmann-scaled profile to better than 1% at every face but the wall face. The remaining error is the extrapolation of the cell-average distribution to the wall face in the first cell, where $Y \propto r^2 e^{-\chi(r)}$ varies strongly. It is first order in the wall-cell size. On uniform octrees, the collected current is 0.71, 0.93, and 0.98 of the exact value at levels 5, 6, and 7 (wall cells 0.28, 0.14, and 0.07 $R_p$). Mesh grading reduces this error. Following Laframboise's computation net [49], the radial axis of each octree is a mapped coordinate (Section 2.6). On the logarithmic map, the level-5 wall cell equals that of a uniform level-7 octree (0.075 $R_p$) at one sixty-fourth of the cells and with the same time step (set by the pitch-angle advection). The collected current becomes 0.92, 0.98 and 0.995 of the exact value at levels 5, 6 and 7. The deficit depends only on the wall cell, regardless of which mesh family supplies it. The graded level-5 mesh matches the uniform level-6 one, and the graded level-6 mesh is within 2% of the Liouville value.

## 4.4. Kinetic versus fluid ions around the sphere

With the ion model switch, we compare orbital kinetics against the radial two-moment fluid model on an identical footing. Both runs use the same electron solver, Poisson shells and boundary conditions, charge-exchange operator, and reservoir; only the ion description differs. Charge exchange destroys the centrifugal barrier that was reflecting ions and converts reflected orbits into collected ones. However, it also removes the sheath-accelerated ions that carry the collisionless current and replaces them with cold ones born at rest. The comparison measures which of the two effects wins. The radial fluid closure has no orbital motion to lose. In addition, it meets the classical sonic-point difficulty of the radial-motion (ABR) limit [59].

The $\lambda_i$ series for the demonstration case ($M/m = 100$, $T_i = T_e$, $R_p/\lambda_{De} = 7.9$, $R_{out} = 3R_p$ with the $r^{-2}$ tail, each run started from the MoC collisionless floating state and taken to the current-balance gate) shows that both branches deepen monotonically with collisionality and that each connects to its own collisionless limit. Over the decade in $\lambda_i/R_p$, the kinetic float deepens by 1.06 $T_e$ and the fluid float by 0.84 $T_e$, while the collected ion current falls by a factor of 2.86 in the kinetic model and 2.32 in the fluid one. The two descriptions differ by a nearly constant offset. The fluid wall floats 0.39 $T_e$ shallower than the kinetic one in the collisionless limit and 0.53 $T_e$ shallower at $\lambda_i = 0.3R_p$. The reason is that the radial fluid, with no angular momentum to carry ions past the sphere, over-collects by 50% without collisions and still by 69% at the collisional end. The electron flux that has to balance the ion current is exponential in the barrier, so a factor-of-three change in the ion supply registers as a one-$T_e$ shift of the wall. Therefore, the collected current, and not the floating potential, remains the discriminating measurement between the two ion descriptions, and the orbital deficit of the fluid model is a constant of the comparison rather than a collisionless peculiarity. In the planar CCP (Section 3.6), where the ions are born at rest inside the discharge and carry no angular momentum, the same fluid closure is 23% low in the bulk. There, the deficit arises through the isothermal closure at the sheath edge and not through orbital motion.

## 4.5. Ion focusing and trapping shape the potential

Two ion populations that standard models omit control the potential profile around an absorbing sphere. Both are geometric consequences of the effective radial potential

$$U_\ell(r) = q\phi(r) + M\frac{\ell^2}{2r^2} \tag{15}$$

For finite angular momentum, the attractive electrostatic well and the centrifugal wall combine into a well plus barrier. The top of the barrier is an X-point of the radial phase plane, and the separatrix through it encloses an island of trapped orbits [57,60,61]. Collisionless orbits from the reservoir cannot enter the island, which is populated and drained only by collisions. A stationary solver that traces characteristics back to a boundary (our MoC solver included) leaves it empty by convention; this is the ion hole of Al'pert et al. [57] and Laframboise [49]. The cost of this convention for stationary Vlasov–Poisson solvers is analyzed in [62]. The classification of the orbits, the reduced two-cone model of Section 2.5 that inverts it analytically, and the ion hole are developed in the companion paper [48]. Here we show only what the transient kinetic solver adds. With collisions and the history of the self-consistent field included, the trapped population is no longer a convention but a result.

The kinetic $\lambda_i$ sweep at fixed bias ($\chi_p = 5$, $R_p/\lambda_{De} = 7.9$, $T_i = T_e$, $R_{out} = 4.5R_p$; [48]) shows the trapped population in the field. The collected ion current falls from 3.35 $\Gamma_{th,i}$ without collisions ($-1.1\%$ from the MoC value 3.385) to 3.17, 2.67 and 1.71 at $\lambda_i/R_p = 10$, 3 and 1, a factor of 1.96 between the collisionless and the $\lambda_i = R_p$ runs. This is because charge exchange removes the sheath-accelerated ions. The potential outside the sheath responds differently. At $\lambda_i = 3R_p$ and $1R_p$, it dips 0.010 and 0.047 $T_e$ below the reservoir level. This is due to the space charge of the charge-exchange-born ions accumulating on bound orbits in the presheath (the filled ion hole). The sweep itself and its comparison with the reduced model and its trapped-bath variable are in [48].

## 4.6. Per-cell selection of the electron solver

Here we apply cell-by-cell selection to the electrons around the sphere. We divide the electron phase space by the total energy $\mathcal{E} = mv^2/2 + \chi(r)T_e$ at the wall barrier $\varepsilon_w = \chi_p T_e$. Electrons below the barrier never reach the surface. Elastic collisions conserve their energy, so their steady distribution is the isotropic reservoir value $f_\infty(\mathcal{E})$ at any collisionality. Electrons above the barrier are absorbed when their orbit reaches the surface. The loss cone empties the outgoing hemisphere next to the wall, and scattering refills it within a few mean free paths (Figure 15a). Outside the loss cone, the distribution is nearly isotropic, and $f_0(r, \mathcal{E})$ obeys the equation

$$v \frac{\partial f_0}{\partial t} = \frac{1}{r^2} \frac{\partial}{\partial r}\left(r^2 v \mathcal{D} \frac{\partial f_0}{\partial r}\right), \tag{16}$$

where $\mathcal{D} = v^2/3\nu_m$, and $v = \sqrt{2(\mathcal{E} - \chi(r)T_e)/m}$ is the speed of an electron of total energy $\mathcal{E}$ at radius $r$. The weight $v$ is the density of states per unit energy, which varies with $r$ at fixed $\mathcal{E}$. For the trapped electrons, the steady-state solution of Eq. (16) is $f_\infty(\mathcal{E})$. Their description thus collapses from the two variables $r$ and $\mathcal{E}$ to the total energy alone, at any collisionality. This is the lowest level of the electron hierarchy. It needs no solver, and the hybrid uses it analytically.

For the escaping electrons, the steady solution is $f_0 = f_\infty(\mathcal{E}) + I(\mathcal{E})G(r, \mathcal{E})$, with $G = \int_r^{R_{out}} 3\,\nu_m ds/(16\pi^2 s^2 v^3)$ and $I$ the inward current of the group. With energy-changing collisions, Eq. (16) acquires the energy-diffusion and inelastic terms of the nonlocal approach [30,31].

*The hybrid solver.* The hybrid electron solver uses the 1d2v octree between the wall and an interface radius $r_i$ and the Fokker–Planck model beyond it. The division is thus made twice, in energy and in radius. In energy, it is made at the wall barrier, between trapped electrons that are known analytically and escaping electrons that are solved for. In radius, it separates the loss-cone layer, where the escaping distribution is anisotropic and the 1d2v octree is kept, from the outer region, where the distribution is nearly isotropic and the Fokker–Planck model in position and total energy suffices. The interface is a radial surface spanning the whole energy range of the escaping electrons. It is not the loss-cone boundary in velocity space, because the loss cone itself lies inside the layer and is refilled there by scattering. The sensor finds the radius beyond which the refilling is complete. In the all-kinetic solution at $\lambda = 0.3R_p$, the anisotropy $|f_1|/f_0$ of the electrons one $T_e$ above the barrier falls from 1.1 at the wall to 0.11 at $1.6R_p$ and 0.03 at the interface (Figure 15b). The anisotropy sensor sets the interface. Namely, $r_i$ is the radius beyond which $\delta_A < \delta$ for every escaping group of the kinetic solution, plus two radial columns as a buffer. However, it is never closer to the wall than $R_p + 3\lambda$, because within a few mean free paths of an absorbing wall the distribution is anisotropic, however collisional the plasma is. Beyond $r_i$, the octree is masked, so the kinetic tree holds only the layer and its moments, which only the sensor and the output need, and gathers them only when evaluated. In the runs below, we fix the interface at the radius selected by the sensor in a preceding run with the full octree, because a Basilisk mask cannot be undone. An interface that moves with the solution requires rebuilding the tree from a leaf checkpoint when the ownership changes.

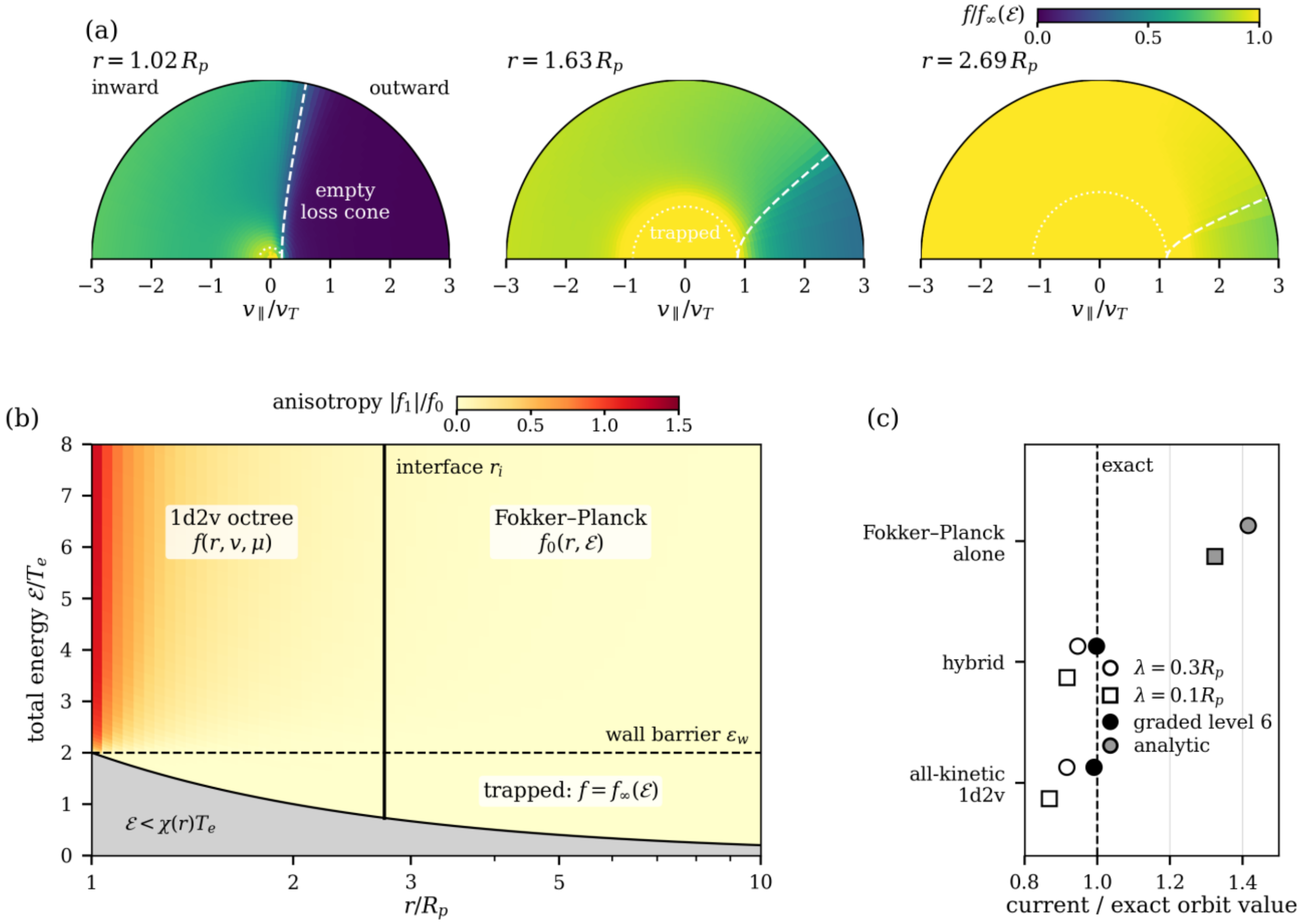

Figure 15. The loss cone and the electron solvers around the sphere ($\chi_p = 2$, prescribed Coulomb potential, $\lambda = 0.3R_p$, all-kinetic run on the graded level-6 mesh). (a) The distribution $f/f_\infty(\mathcal{E})$ in the plane of the velocity components along and across the radius at three radii, outward to the right; dashed: the collisionless loss-cone boundary; dotted: the speed below which an electron cannot reach the wall (trapped electrons). (b) The anisotropy $|f_1|/f_0$ of the same solution (first over zeroth Legendre moment in $\mu$) in the plane of radius and total energy, with the regions of the hybrid: the 1d2v octree inside the interface $r_i = 2.74R_p$, the Fokker–Planck model beyond it, and the trapped electrons below the wall barrier $\varepsilon_w$, known analytically. (c) The collected current over the exact orbit value for the Fokker–Planck model alone (analytic), the hybrid at $\delta = 0.05$ and the all-kinetic solver, at $\lambda = 0.3R_p$ (circles) and $0.1R_p$ (squares); open symbols: graded level-5 mesh; filled: graded level-6 mesh.

*Conservation.* The two models are coupled per velocity row of the octree at $r_i$, that is, per energy group. The octree supplies the outward partial current $J_+$ of each row through the interface face, tallied from the face fluxes its own transport uses. The Fokker–Planck model returns the inward partial current $J_-$ (Marshak's half-range moment of the P1 distribution $f = f_0 - \mu\lambda\partial f_0/\partial r$ with $\lambda = v/\nu_m$). In the discrete row, this current is linear in the group current, $J_- = P + QI$. The balance between the net current of the row and $I$ fixes $I$, and with it $f_0$, in closed form. The masked ghost cells beyond $r_i$ carry the P1 state, so the octree receives the inflow with its angular slope through ordinary upwind advection. By construction, the net current of the octree through the interface equals the current of the Fokker–Planck region in every row, and the only particle reservoir is the Maxwellian at $R_{out}$, as in the all-kinetic run. Every row is coupled in this way, including the rows of the trapped electrons. Masking the trapped region instead, with either a reflecting (Liouville) or a zero-flux edge at $\varepsilon = \varepsilon_w$, fails once the electrons collide. The barrier surface cuts the octree cells in a staircase, and the two edges injected 1–12% of the collected current and biased it by +15%, respectively.

*Results.* Table 2 gives the results for the repelled electrons of Section 4.3 ($\chi_p = 2$, prescribed Coulomb potential, $R_{out} = 10R_p$, graded level-5 mesh) with isotropic elastic scattering at constant $\nu_m$, at mean free paths $\lambda = v_T/\nu_m = 0.3R_p$ and $0.1R_p$ and thresholds $\delta = 0.1$, 0.05 and 0.02. We start each run from the exact solution of the same problem obtained by integration along the orbits [48], which gives $i\, e^{\chi_p} = 0.609$ and 0.305 and against which every part of the hybrid can be audited, and run it for 60 000 steps ($t = 295R_p/u_0$). In all six hybrids, the net current of the octree through the interface equals the Fokker–Planck current to the four digits printed by the audit. In the layers of $\delta = 0.1$ and 0.05, the current is the same from the wall face to the interface face to four or five digits, so these runs are steady and conservative. The thickest layers ($\delta = 0.02$, reaching $4.2R_p$ and $2.6R_p$) were still filling, because their diffusive relaxation time $(r_i - R_p)^2/D$ is comparable with the length of the run. In the all-kinetic run, the current through the outer faces at $\lambda = 0.3R_p$ ranges from 4% below to 15% above its wall value beyond $3R_p$. This is due to the slow diffusive filling of the coarse outer cells, which the hybrid replaces by a quadrature. The last two rows of Table 2 (SC) repeat the pair at $\lambda = 0.3R_p$, $\delta = 0.05$ with the potential computed self-consistently: Poisson's equation on the field mesh, kinetic ions as in Section 4.2, and the wall held at $\chi_p = 2$. The hybrid reads the potential from the field solver at every step, so its wall barrier and the potential energy of the Fokker–Planck region follow the solution. It recomputes the interface coefficients every 200 steps. Both runs start cold and run 120 000 steps ($t = 524R_p/u_0$), the all-kinetic one twice as long, because the density in its layer was still relaxing at the end of the first segment. The interface audit closes to four digits as before.

*Accuracy.* On the level-5 mesh the hybrid current exceeds the all-kinetic current by 2 to 4% at $\lambda = 0.3R_p$ and by 4 to 7% at $0.1R_p$, the difference decreasing with the threshold (Table 2 and Figure 15c). With the self-consistent potential (Table 2, SC), the excess is the same as with the prescribed potential at the same threshold and interface. The ion current, the potential and the density beyond the interface agree to better than 0.5%, and the density in the layer to 6%. No orbit solution exists for this case, so the last column is empty. Both runs carry the wall-cell deficit of Section 4.3. Divided by its collisionless value, the all-kinetic current at $\lambda = 0.3R_p$ is 0.609 against the exact 0.608, so its deviation from the exact current is due to the wall cell alone. The graded level-6 mesh separates the two errors (Table 2, $\lambda = 0.3R_p$, $\delta = 0.05$, the same interface radius). There, the all-kinetic current is 0.9% below the exact value, and the hybrid current is 0.2% below it. At a fixed threshold and interface, the difference between the two models thus falls from 3.3% to 0.7% when the mesh is refined. Since the error of the P1 closure at the interface does not depend on the mesh, it is at most 0.7% here, and the rest was a discretization error of the coupling on the coarse velocity rows. The hybrid converges to the exact current with the mesh, as the all-kinetic solver does. The Fokker–Planck model alone marks the other end of the ladder. Carried to the wall (with the Marshak condition of the coupling applied at the absorbing surface instead of at $r_i$), it gives $ie^{\chi_p} = 0.86$ at $\lambda = 0.3R_p$ and 0.40 at $0.1R_p$, 42 and 32% above the exact currents. In the collisionless limit, it gives twice the Liouville value, since the P1 closure cannot represent the half-empty loss cone at all. The error of the reduced model is concentrated in the loss-cone layer, which the kinetic layer of the hybrid supplies. Figure 15c compares the three levels. At $\lambda = 0.1R_p$, the all-kinetic level-5 run is not steady, and its deviation includes a mean free path not much larger than the graded wall cell ($0.075R_p$).

*Cost.* Removing the octree beyond the interface reduces the electron tree 1.6 to 5.3 times and the CPU time 1.2 to 3.3 times (Table 2). Across the six hybrids, the CPU time is linear in the number of octree cells (1.2 s per 1000 cells and 1000 steps plus a fixed 2.7 s, against 1.05 s per 1000 cells for the all-kinetic run). The gain grows as the layer becomes thinner, that is, with the collisionality and with the threshold. On the level-6 mesh, the same interface removes 56% of the cells and 39% of the CPU time. In the self-consistent pair, whose time includes the ion octree and the field solve, it removes 56% of the electron cells and 11% of the total. The fixed part evaluates the Fokker–Planck solution on the field cells and the P1 state of the ghost cells at every step. The per-cell excess is the masked part of the tree, which Basilisk still traverses, and a root box ending at the interface would remove it. In 1d2v on the logarithmic radial map, the Fokker–Planck region replaces only the coarse outer part of the octree. The economy is therefore modest here, and it grows with domain size, collisionality, and phase-space dimension. The construction carries over to the ions, whose reduced model inside a fluid region is the 1d1v(S) solver of Section 2.5; that hybrid is the subject of [48].

Table 2. Electron hybrid for the repelled electrons of Section 4.3 ($\chi_p = 2$, Coulomb potential, $R_{out} = 10R_p$, graded level-5 mesh; L6: graded level-6 mesh; SC: self-consistent potential with kinetic ions, CPU time including the ion and field solvers) with elastic scattering at constant $\nu_m$, after 60 000 steps (50 000 on level 6, 120 000 with SC): interface radius, octree cells, CPU time per 1000 steps on one core of a shared workstation, the collected current, and its deviation from the all-kinetic run and from the exact orbit solution (0.609 and 0.305).

| $\lambda/R_p$ | $\delta$ | $r_i/R_p$ | cells | CPU (s) | $i\,e^{\chi_p}$ | vs all-kinetic | vs exact |
|---|---|---|---|---|---|---|---|
| 0.3 | all-kinetic | – | 32 768 | 34.3 | 0.558 | – | −8.4% |
| 0.3 | 0.1 | 2.21 | 11 264 | 16.3 | 0.583 | +4.4% | −4.3% |
| 0.3 | 0.05 | 2.74 | 14 336 | 19.6 | 0.576 | +3.3% | −5.4% |
| 0.3 | 0.02 | 4.22 | 20 480 | 27.9 | 0.569 | +2.0% | −6.5% |
| 0.1 | all-kinetic | – | 32 768 | 34.5 | 0.265 | – | −13.2% |
| 0.1 | 0.1 | 1.54 | 6 144 | 10.4 | 0.284 | +7.2% | −6.9% |
| 0.1 | 0.05 | 1.91 | 9 216 | 13.9 | 0.280 | +5.5% | −8.4% |
| 0.1 | 0.02 | 2.55 | 13 312 | 18.7 | 0.276 | +4.3% | −9.4% |
| 0.3, L6 | all-kinetic | – | 262 144 | 244.9 | 0.603 | – | −0.9% |
| 0.3, L6 | 0.05 | 2.74 | 114 688 | 150.5 | 0.608 | +0.7% | −0.2% |
| 0.3, SC | all-kinetic | – | 32 768 | 148 | 0.559 | – | – |
| 0.3, SC | 0.05 | 2.74 | 14 336 | 132 | 0.578 | +3.4% | – |

## 5. Discussion and outlook

Building on prior work, we conceived the PAMAR architecture for adaptive kinetic-fluid plasma simulations. We have developed and demonstrated the capability to use separate adaptive Cartesian meshes for ions, electrons, and the electric field in two settings: transient CCP simulations with 1d1v kinetic solvers for electrons and ions, and 1d2v simulations of spherical plasmas with explicit time marching. We identified three levels of charged-particle transport in collisional low-temperature plasma: kinetic, angular moments with closures (an intermediate kinetic level), and fluid (velocity moments).

The PAMAR solvers (electric field, electron, and ion) currently run sequentially in a single process. CPU parallelism is achievable at two levels: intra-solver thread parallelism with OpenMP across phase-space grid cells, and inter-solver concurrency by advancing the solvers simultaneously between field solves (the solver interface already permits this). Distributed-memory MPI scaling will require the moment-integration functions to include global reductions. After that, Basilisk's existing tree-partitioning infrastructure along space-filling curves should handle domain decomposition for each species solver transparently. The lessons of the UFS port to heterogeneous systems [14] apply directly. The load must be rebalanced after every adaptation, and a kinetic cell costs orders of magnitude more than a fluid one.

The GPU path of PAMAR is constrained by Basilisk's current GPU backend, which supports only Cartesian grids and not the adaptive quadtrees and octrees used by the kinetic solvers. Switching to uniform phase-space grids would enable transparent GPU execution of the advection and collision kernels through Basilisk's existing GLSL compute-shader infrastructure. The price is the adaptive economy measured in Section 4.1. The split (AMPS) formulation of [8,9], in which each configuration-space cell owns a uniform or mildly adaptive velocity grid, is the natural intermediate choice, and the CUDA kernels of [21,22] were written for it.

Electron solvers impose the most severe time-step restriction in PAMAR. In transient CCP runs, the time step is $10^{-4}$ of the RF period and is limited by electrons. Long simulations are required to reach a time-periodic steady state controlled by ambipolar ion diffusion. In the spherical runs at prescribed bias, we decouple the clocks to reach the steady state more quickly and scale the mass ratio. Replacing explicit time advance with an implicit scheme for electrons coupled to Poisson's solver would let the coupled system advance on the ion or RF time scale, reducing the cost by several orders of magnitude. The explicit ion and neutral solvers would remain unchanged to balance memory demands for multi-component plasmas.

The hybrid ion solver of Section 3.7 also shows where the cost of a transient CCP run goes. In the reference case, the ion solver takes about 40% of the run time and the electrons the rest. Removing the kinetic ion tree from the fluid region (a fifth of the gap) would save a few percent. Because a Basilisk mask cannot be undone while the interface moves, this requires either rebuilding a tree from a checkpoint when ownership changes or using two kinetic ion solvers confined to fixed windows near the electrodes. A larger saving comes from the ion time step. The kinetic ions are advanced at the coupled step set by the electrons, about $1.6 \times 10^{-4}$ of the RF period, although their own stability limit is 0.03 of the period. Advancing them, with their adaptation, only on their own time scale (about 65 times per period, with the field and the ionization source averaged over the ion step) reduced the CPU time per RF period by a third in a 10-period test of the reference case, with changes below 1% in the densities, the potential, the ion flux and the ion energy distribution at the electrodes. The electron time step remains the main limitation, and the implicit electron solver discussed above is the main target.

Per-cell selection addresses spatial nonlocality, and the same machinery addresses temporal nonlocality. Because the hybrid of Section 3.7 carries both ion descriptions in every cell, the switch between them can be driven by time as well as by position. A run can be carried to its periodic state with the fluid solver and then handed to the kinetic solver for the last few RF periods, everywhere or where a criterion asks for it. A transient event such as breakdown can be computed kinetically even in a collisional regime in which the static criteria would select the fluid model, with the fluid solver taking over again once the event has passed. We tested this switch in time on the reference case (Figure 16). The run starts from the converged fluid solution of Section 3.6, with the kinetic ions seeded with the fluid moments, and after five fluid periods, every cell is handed to the kinetic solver. The switch is smooth: the center density dips by 10% over the first 20 periods, while the seeded distributions are replaced by the kinetic ones, and returns to the fluid value by period 60 (Figure 16a). The quantities set by the sheath reach their kinetic values long before the bulk does. The ion velocity at the electrodes falls from the fluid value, 5.52 $v_s$, to the kinetic one, 5.26 $v_s$, in the first kinetic period and stays within a few percent of it (Figure 16b). A hundred periods after the switch, the ion energy distribution at the electrodes matches the converged kinetic run (Figure 16c), with the same peak at 115 eV, whereas the center density has risen only from 0.126 to 0.129 $n_0$ and is still 21% below the kinetic value. The particle balance of the discharge sets the bulk density, which relaxes over hundreds of periods (Section 3.2). A kinetic finish of a few periods therefore corrects the electrode quantities of a fluid run but not its bulk density.

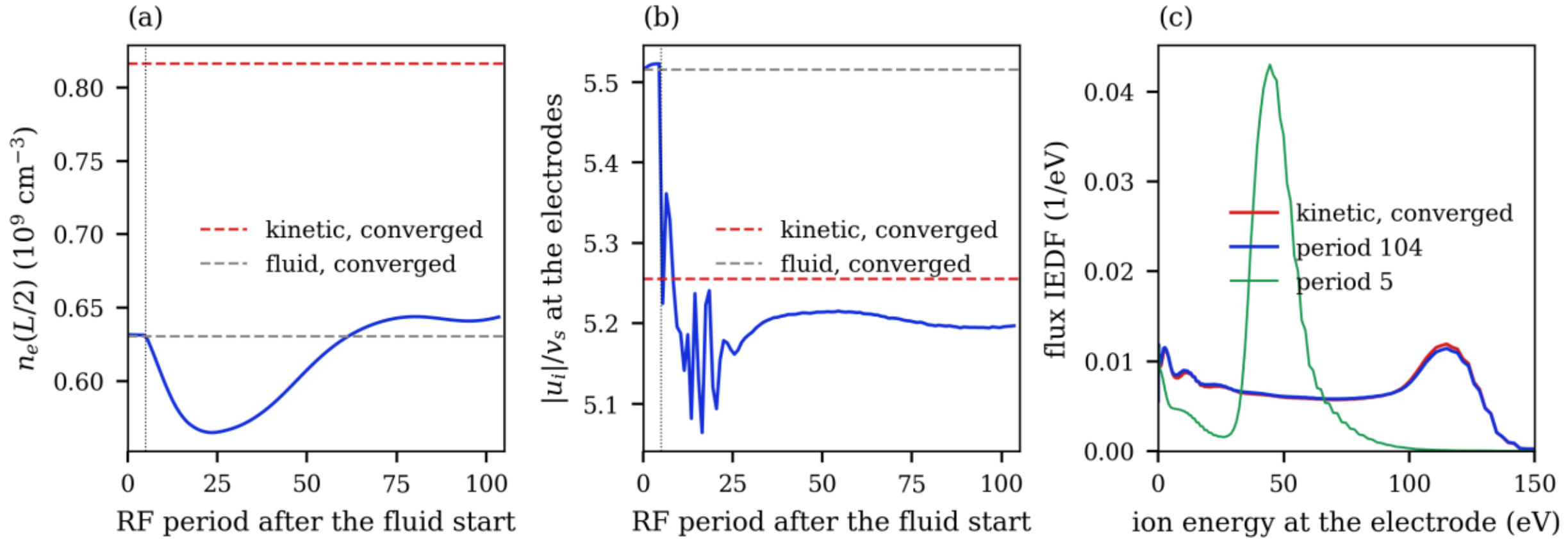

Figure 16. Switching the ion solver in time: the reference case started from the converged fluid solution of Section 3.6 and was handed to the kinetic solver at period 5 (dotted line). (a) Center electron density against the converged fluid and kinetic values. (b) Ion drift velocity at the electrodes (0.05 cm from each, mean of the two sides, period-averaged) against the same two values. (c) Flux-weighted ion energy distribution at the right electrode during the first kinetic period and 100 periods after the switch, against the converged kinetic run.

The CCP demonstration showed that the body of the EEDF depends on total energy, while the tail is temporally nonlocal. A homogenized Fokker–Planck description in $(x, \mathcal{E})$ for the body [19,20,31], supplemented by an angular-moment description of the fast, anisotropic group, would capture the discharge at a fraction of the kinetic cost. The kinetic solver would then serve as the benchmark for calibration and validation. The electron groups observed in CCP resemble those in the cathode region of DC discharges [6,63]. The angular-moment level is the right tool for efficiently describing electrons in gas-discharge plasmas. The two-stream model of Section 3 and the Fokker–Planck model of Section 4.6 are its first two members in PAMAR. Developing the general angular-moment solver, with the kinetic solver run alongside to show where it holds, is the next step for the electron side of PAMAR.

Extending the technology to problems such as laser ablation into a background gas, where solid, melt, and plasma/gas regions are each governed by their own equation set, requires coupling independent solver instances across material regions that exchange fluxes and boundary conditions at moving interfaces tracked by level-set or volume-of-fluid methods already available and validated in the framework [17]. A simpler form of this coupling between independent solver instances is already used in our adaptive-Cartesian-mesh simulations of laser-generated plasmas [64,65], where a gas-dynamic solver for the gas phase, a solver for the metal and molten pool, and a solver for laser power propagation run as three coupled instances. The coupling mechanism described here, in which independently compiled solvers with localized symbols coexist in one executable, enables this.

The present demonstrations use phase spaces of dimension $\leq 3$. For higher phase-space dimensions, we plan to split configuration and velocity space, as previously done in UFS [8].

## 6. Conclusions

We have described PAMAR, which extends the Adaptive Mesh and Algorithm Refinement methodology to weakly ionized plasma by making each species an independent adaptive-mesh solver with its own phase-space dimensionality and refinement criteria, coupled through a field solver in one executable. We have demonstrated this capability for a planar capacitive discharge computed with 1d1v two-stream kinetic solvers and for a floating sphere in a weakly collisional plasma computed with 1d2v octree solvers. In both cases, a Poisson solver coordinated the run, selecting and driving the transport models for both species. We demonstrated a hierarchy of transport models for charged species and developed multi-species solvers ranging from fully kinetic to fluid, selectable globally or cell by cell. A hybrid ion solver keeps the kinetic solver where its own solution shows the ion temperature above 1.5 times the gas temperature and uses the fluid model elsewhere. It reproduces kinetic ions to 1.5% with a fifth of the gap fluid. A hybrid electron solver keeps the 1d2v octree only for the escaping electrons in the loss-cone layer and uses a Fokker–Planck model in position and total energy elsewhere. Both hybrids use the reduced description (the two-moment fluid model for the ions and the Fokker–Planck model in position and total energy for the electrons) wherever the kinetic solution, computed alongside, shows that it holds.

The CCP results resolve, without the statistical noise of PIC and with RF-phase resolution, the dynamic properties of CCP:  the formation of electron groups, the two time-averages of the potential that govern electrons and ions, the secondary peaks of the ion energy distribution, and the transition at 5 MHz to a regime in which the sheath oscillation spans the gap. They show the temporal modulation of the EEDF tail, help clarify the validity limit of fast CCP models, and offer insights complementary to PIC.

The spherical results show that the adaptive ion octree resolves the cold ions at a tenth of the uniform-mesh cost, and that graded meshes improve absorbed flux, collected currents, and density profiles. Per-species clocks accelerate the approach to the steady state. Fluid ions over-collect by 50–70% in spherical geometry because they carry no angular momentum. In the planar discharge, where ions are born at rest and carry no angular momentum, their error comes from the isothermal closure at the sheath edge, and the per-cell selection in Section 3.7 confines them to the bulk, where the closure holds.

Future work will focus on parallelizing the species solvers, coupling an implicit electron solver to Poisson's equation, developing a split phase-space formulation for multi-dimensional and highly transient problems, and coupling independent solver instances across material regions for multi-phase plasmas.

## Acknowledgments

This work was supported by the NSF EPSCoR FTPP project (Cooperative Agreement OIA-2148653).

## Declaration of competing interests

The authors declare no known competing financial interests or personal relationships that could have influenced the work reported in this paper.

## Data availability

Data will be made available on request.

## Declaration of generative AI and AI-assisted technologies in the manuscript preparation process

During this work, the authors used Claude Code (Anthropic) to assist with code development, data analysis, and manuscript editing. After using this tool, the authors reviewed and edited the content as needed and take full responsibility for the published article.